\documentclass[12pt]{article}

\renewcommand{\i}{\mathrm{i}}
\renewcommand{\d}{\mathrm{d}}

\newcommand{\be}{\begin{equation}}
\newcommand{\ee}{\end{equation}}
\renewcommand{\tilde}{\widetilde}
\renewcommand{\i}{\mathrm{i}}
\renewcommand{\d}{\mathrm{d}}

\usepackage{jheppub}

\usepackage{subcaption}
\usepackage{xcolor}
\usepackage{amsmath}
\usepackage{amssymb}
\usepackage{amscd}
\usepackage{dsfont}
\usepackage{enumerate}
\usepackage{enumitem}
\usepackage{amsfonts}
\usepackage{epsfig}
\usepackage{mathtools}
\usepackage{yfonts}
\usepackage{bbold}
\usepackage{breqn}
\usepackage[utf8]{inputenc}
\usepackage[english]{babel}
\usepackage{caption}
\usepackage{multirow}

\usepackage{titlesec}

\titleformat{\paragraph}
{\normalfont\bfseries}
{}{0em}{}
\titlespacing*{\paragraph}
{0pt}{3.25ex}{1.5ex} 

\renewcommand{\tilde}{\widetilde}
\renewcommand{\i}{\mathrm{i}}
\renewcommand{\d}{\mathrm{d}}

\def\eqa{\begin{eqnarray}}
\def\eqae{\end{eqnarray}}
\def\eq{\begin{equation}}
\def\eqe{\end{equation}}
\def\be{\begin{equation}}
\def\ee{\end{equation}}
\def\bea{\begin{eqnarray}}
\def\eea{\end{eqnarray}}
\def\ba{\begin{array}}
\def\ea{\end{array}}

\title{A semiclassical ramp from RT surfaces}
\author[a,c]{Huajia Wang,}
\author[b,c]{Zhenbin Yang}
\affiliation[a]{School of Quantum, Kavli Institute for Theoretical Sciences, University of Chinese Academy of Sciences, Beijing 100190, China}
\affiliation[b]{Chen-Ning Yang Institute for Advanced Study, Tsinghua University}
\affiliation[c]{Peng Huanwu Center for Fundamental Theory, Hefei, Anhui 230026, China}
\emailAdd{wanghuajia@ucas.ac.cn}
\emailAdd{yangiastu@gmail.com}

\abstract
{We study the ramp in the modular spectral form factor (SFF) associated with a subregion of a semiclassical state in AdS/CFT. In cases of spherical regions in a vacuum state, the modular SFF reduces to the thermal SFF of hyperbolic black holes, with the ramp arising from the associated double-cone geometry. In AdS$_3$/CFT$_2$, this double-cone geometry is an analytic continuation of the annulus wormhole.
We further consider two distant spherical sub-regions as a perturbation of the modular $U(1)$-symmetric setting. The modular SFF exhibits initial $T^2$ growth from two double-cones. We study how correlations between them give rise to a single linear ramp at late times. }

\begin{document}
\maketitle

\section{Introduction}
The relation between black holes and quantum chaos has become a central theme in the study of quantum gravity. In AdS/CFT \cite{Maldacena:1997re,Gubser:1998bc,Witten:1998qj}, it provides a way to study near horizon dynamics of black holes through universal properties of the boundary theory including information scrambling \cite{Hayden:2007cs,Sekino:2008he,Shenker:2013pqa,Maldacena:2015waa} and the development of random matrix statistics \cite{Cotler:2016fpe,Saad:2018bqo,Saad:2019lba}.
Among these, the double-cone wormhole serves as a basic example of a universal relation between black hole horizons and random matrix behavior \cite{Saad:2018bqo,Chen:2023dc}. Its construction applies to any eternal two-sided black hole geometry with a bifurcate Killing horizon.

This connection is expected to extend to more general Ryu--Takayanagi (RT) surfaces. The starting point is the analogy between black hole thermodynamics and holographic entanglement. For a boundary sub-region $A$ in a holographic state $\psi$, the RT formula \cite{Ryu:2006bv,Ryu:2006ef} gives
\be\label{eq:RT}
S_A=-\operatorname{Tr}\rho_A\ln\rho_A
=\frac{\operatorname{Area}(\Sigma_A)}{4G_N}+\cdots,
\ee
where $\rho_A=\operatorname{Tr}_{H_{\bar A}}|\psi\rangle\langle\psi|$,\footnote{We use a UV regulator so that reduced density matrices and their spectra are well-defined, setting aside the type-III algebraic subtleties of continuum QFT \cite{Witten:2018zxz}.} and the omitted terms include bulk quantum corrections \cite{Faulkner:2013ana}.
The RT surface $\Sigma_A$ generalizes the role of the horizon in the Bekenstein--Hawking entropy formula \cite{Bekenstein:1973ur,Hawking:1974sw}. Correspondingly, the modular Hamiltonian
\be
K_A=-\ln\rho_A
\ee
generalizes the thermal Hamiltonian, with modular flow $O(s)=e^{iK_A s}Oe^{-iK_A s}$ replacing physical time evolution. This suggests extending the horizon-based probes of chaos to the entanglement structure of general holographic states, including states whose bulk duals contain no black hole. Related directions include modular-flowed correlators, OTOC-inspired bounds on modular scrambling, and connections between modular evolution and the quantum ergodic hierarchy \cite{Faulkner:2018faa,Jan:2020,ChandrasekaranFaulknerLevine,Ouseph:2023juq,deBoerEtAl}. 

In this work, we spell out some details of the generalization of the double-cone wormhole to more general RT surfaces by studying the modular SFF,
\be\label{eqn:mSFF}
F_A(\beta,T)=\overline{\left(\operatorname{Tr}e^{-(\beta+iT)K_A}\right)
\left(\operatorname{Tr}e^{-(\beta-iT)K_A}\right)},
\ee
where $T$ is modular time and $\beta$ is a modular inverse temperature; the two traces are R\'enyi moments continued to $\beta\pm iT$.\footnote{Entanglement spectral correlations have been investigated in many-body wavefunctions \cite{ChenLudwig}, quantum circuits \cite{ChangEtAl}, and Haar random states with a fixed-area gravitational realization \cite{CaoFaulkner}.} Our starting point is to view \eqref{eqn:mSFF} as a single trace in the doubled Hilbert space of two CFTs, and consider the contribution from the boost invariant states such as the  canonical purification of $\rho_A$ and its one-sided modular flow. In the bulk, the modular flow generates a kink transformation at the RT surface \cite{Bousso:2020yxi,Kaplan:2022orm} together with modular evolution of the matter fields \cite{JLMS}. We propose extending the double-cone wormhole by a modified sum over these boost invariant states where the summation includes a compactification of the relative modular evolution time with periodicity $T$. This is the same modified trace on the double cone wormhole when we sum over the TFD states \cite{Chen:2023dc}. 

When $A$ is a spherical sub-region in a vacuum state, $\rho_A$ admits a $U(1)$ family of conformal isometries fixing $\partial A$, and the RT surface is the fixed locus of the bulk Killing flow. The proposed construction reduces to the double cone wormhole for a hyperbolic black hole via a conformal transformation that maps the reduced density matrix to a thermal state on hyperbolic space \cite{Casini:2011kv}.  In the case of AdS$_3$ that we studied in detail, this is essentially the annulus wormhole studied in \cite{Jafferis:2025yxt,Jafferis:2025jle}.

We don't have a generic study of our proposal for generic sub-regions that lack U(1) isometry due to the complication when the modular flow is non-local. The only case we have is a perturbative study of the breaking of U(1) isometry from bulk matter correlations, in the case of a pair of distant spherical regions $A_1$ and $A_2$ in the CFT vacuum.  In this case, the classical RT surface and entanglement wedge are disconnected. Neglecting correlations between the regions gives a sum of two decoupled modular Hamiltonians. Each sphere then admits a hyperbolic black hole description and a double-cone contribution, so their combined modular SFF has a contribution proportional to $T^2$. Bulk quantum correlations couple the two modular systems and break the symmetry under independent modular flows. These correlations generate a diffusive coupling whose integrated action grows with modular time and lifts the antisymmetric collective zero modes of the two double cones. This freezes the antisymmetric sector while preserving the symmetric sector that supports a single linear ramp. The resulting contribution exhibits a \mbox{$T^2\!\to T$} crossover. This gives a concrete mechanism by which bulk quantum correlations produce the expected single RMT ramp even when the classical entanglement wedge remains disconnected, and a tractable extension of the double-cone picture beyond the case of exact modular $U(1)$ symmetry.

The paper is organized as follows. In section (\ref{sec:modular_SFF}), we set up conventions and discuss the proposed modified sum for the ramp in modular SFF. Section (\ref{sec:geometric_flow}) develops the explicit construction for regions with modular $U(1)$ symmetry, focusing on a single arc in the vacuum of a two-dimensional CFT. Section (\ref{sec:two_spheres}) studies the case of two distant spheres. We conclude in section (\ref{sec:discussion}).

\section{Modular spectral form factor}\label{sec:modular_SFF}
Let $\rho_A$ be the normalized reduced density matrix of a boundary sub-region $A$. On its support, define
\be\label{eqn;KaZa}
K_A=-\ln\rho_A,\qquad Z_A(z)=\operatorname{Tr}\rho_A^z
=\operatorname{Tr}e^{-zK_A}.
\ee
The modular spectral form factor (SFF) is
\be\label{eq:modular_sff_definition}
F_A(\beta,T)=\overline{Z_A(\beta+iT)Z_A(\beta-iT)},
\ee
where $\beta>0$ is a modular inverse temperature and $T$ is modular time. The overline denotes an ensemble average when available, or a suitable smoothing in modular time for a fixed theory. The modular SFF probes correlations between the eigenvalues of $K_A$, and hence the entanglement level statistics of $\rho_A$.

These definitions require a UV prescription \cite{Witten:2018zxz}. We use an entanglement cut-off: in the Euclidean preparation of $\rho_A$, we excise a tube of radius $\delta$ about $\partial A$ and impose a boundary condition on the new boundary $\mathcal{B}$. The boundary condition is part of the regulated definition, as in the boundary-state treatment of entanglement cuts in two-dimensional CFTs \cite{Cardy:1989ir,Hung:2019bnq}. We keep this prescription fixed and work in a range of $\beta$ where the traces converge.

We focus on holographic states with semiclassical duals and choose a regulator boundary condition described by a tensionless bulk brane $\mathcal{N}_{\mathrm{ent}}$ ending on $\mathcal{B}$.  At leading order in $G_N$, its Neumann boundary condition is \cite{Fujita:2011fp}:
\be\label{eq:tensionless_brane_bc}
K_{ab}-K h_{ab}=0,
\qquad\text{equivalently}\qquad K_{ab}=0,
\ee
where $h_{ab}$ and $K_{ab}$ are the induced metric and extrinsic curvature of $\mathcal{N}_{\mathrm{ent}}$, and $K=h^{ab}K_{ab}$. Thus the brane is totally geodesic. 

The behavior of $Z_A(\beta+iT)$ is directly related to the Euclidean R\'enyi calculation. For a positive integer $n$,
\be\label{eq:modular_replica_moment}
Z_A(n)=\operatorname{Tr}\rho_A^n
=\frac{Z[\Sigma_A^n]}{Z[\Sigma_A^1]^n},\qquad
\ln Z_A(n)=(1-n)S_A^n,
\ee
where $\Sigma_A^n$ cyclically glues $n$ copies along $A$, with the same regulator on each copy. Assuming replica symmetry, the bulk fixed surface $\Sigma_A^{(n)}$ depends on $n$ and approaches the RT surface $\Sigma_A$ as $n\to1$ \cite{Lewkowycz:2013nqa}. Continuing this saddle to $n=\beta\pm iT$ gives a disconnected contribution to the SFF. In the case when $\Sigma_A$ is not empty, the corresponding modular spectral density $\nu_A(\mathcal{E})$ will be smooth, and its Fourier transform decays at late times:
\be
Z_{A,\mathrm{smooth}}(\beta+iT)
=\int\d\mathcal{E}\;\nu_A(\mathcal{E})e^{-\beta\mathcal{E}}e^{-iT\mathcal{E}}
\longrightarrow0.
\ee

For a chaotic modular spectrum, after resolving exact symmetry sectors, we expect the modular SFF to exhibit universal random matrix correlations described by a ramp-plateau structure. We now try to identify the semiclassical origin of the ramp.

For this purpose, it is useful to view the SFF as an evolution problem in a doubled Hilbert space. The product of traces in (\ref{eq:modular_sff_definition}) can be written exactly as one trace on
\be
\mathcal{H}_{\mathrm{dbl}}=\mathcal{H}_A\otimes\mathcal{H}_A^*,\qquad
K_A^L=K_A\otimes\mathds{1},\qquad
K_A^R=\mathds{1}\otimes K_A^*.
\ee
Here the star denotes dual/transpose. Then
\be\label{eq:modular_sff_doubled_trace}
F_A(\beta,T)=\overline{\operatorname{Tr}_{\mathcal{H}_{\mathrm{dbl}}}
\left(e^{-\beta(K_A^L+K_A^R)}e^{-iT(K_A^L-K_A^R)}\right)}.
\ee
The generator $K_A^L-K_A^R$ evolves the two copies with opposite orientations, and $e^{-\beta(K_A^L+K_A^R)}$ supplies the spectral weight. The trace closes the evolution after modular time $T$. If there are stationary states under the modular evolution, those will lead to non-decaying contributions to $F_A$. One obvious class consists of the product states $|m\rangle_L|m^*\rangle_R$, where $m$ labels an eigenstate of $K_A$. But these states are not semiclassical. Instead, in order to identify a semiclassical state that leads to a candidate semiclassical contribution, we can consider the TFD states by superposing $|m\rangle_L|m^*\rangle_R$:
\be\label{eq:modular_purification_family}
|\Psi_{\alpha,s}\rangle
=\frac{|\rho_A^{\alpha+is}\rangle}{\sqrt{Z_A(2\alpha)}}
=\frac{1}{\sqrt{Z_A(2\alpha)}}\sum_m
 e^{-(\alpha+is)K_m}|m\rangle_L|m^*\rangle_R.
\ee
Here $\alpha>0$ and $s$ is real, and we use the operator-state identification $\langle X|Y\rangle=\operatorname{Tr}X^\dagger Y$. For a chaotic spectrum, $K_m$ will be continuous in the large $N$ limit, and $\alpha$ and $s$ become classical phase-space variables.

At $s=0$, this is the canonical purification of the reweighted density matrix $\rho_A^{2\alpha}/Z_A(2\alpha)$. Varying $\alpha$ changes its one-sided modular energy,
\be\label{eq:modular_purification_energy}
\mathcal{E}(\alpha)=\langle K_A^L\rangle_{\alpha,s}
=-\frac{1}{2}\partial_\alpha\ln Z_A(2\alpha).
\ee
And varying $s$ generates a one-sided modular flow that keeps the reduced density matrices fixed, while changing the correlations between the two sides. We will review its bulk interpretation below based on JLMS.

All these states are stationary under the doubled evolution.
Keeping only their contributions\footnote{The canonical phase-space measure can be obtained by inserting a resolution of identity in terms of wavepackets labeled by the modular energy \(\mathcal{E}\) and its conjugate shift \(s\). The wavepackets have an energy width \(\delta\), chosen much larger than the level spacing but small compared with macroscopic energy scales. Their normalization ensures that the resulting measure \(d\mathcal{E}\,ds/(2\pi)\) is independent of \(\delta\) in the large-\(N\) limit.}  in \eqref{eq:modular_sff_doubled_trace} then leads to the following integral:
\bea\label{eq:modular_ramp_phase_space}
F_A^{\mathrm{con}}(\beta,T)&\simeq&
{1\over 2\pi}\int\d\mathcal{E}\int\d s\;e^{-2\beta\mathcal{E}}.
\eea
Since the modular energy $\mathcal{E}$ and the one-sided modular parameter $s$ are canonically conjugate variables, we have the canonical phase-space measure $\d\mathcal{E}\,\d s/(2\pi)$. 
This summation diverges due to the $s$ integral. In the case of the double cone wormhole, one encounters the same type of integral when we cut the wormhole along a spatial slice \cite{Chen:2023dc}. In that case, $s$ labels the relative time shift between the two sides, and is periodically identified between $s$ and $s+T$. In the hydrodynamic approach of SFF \cite{winer2022hydrodynamic}, this is related to the compactification of the time reparametrization mode, where the integrals over $\mathcal{E}$ and $s$ are the zero-mode integrals. Under this compactification, the $s$ integral leads to a linear ramp:
\be
{1\over 2\pi}\int\d\mathcal{E}\int_0^T\d s\;e^{-2\beta\mathcal{E}}=\int\d\mathcal{E} e^{-2\beta\mathcal{E}} {T\over 2\pi}.
\ee
The factor $e^{-2\beta\mathcal{E}}$ is the spectral weight from the two traces. An inverse Laplace transform with respect to $\beta$ takes us to the microcanonical ensemble, fixing $\mathcal{E}=E$ and therefore $\alpha$. 

\begin{figure}[htbp]
 \centering
 \includegraphics[width=0.90\textwidth]{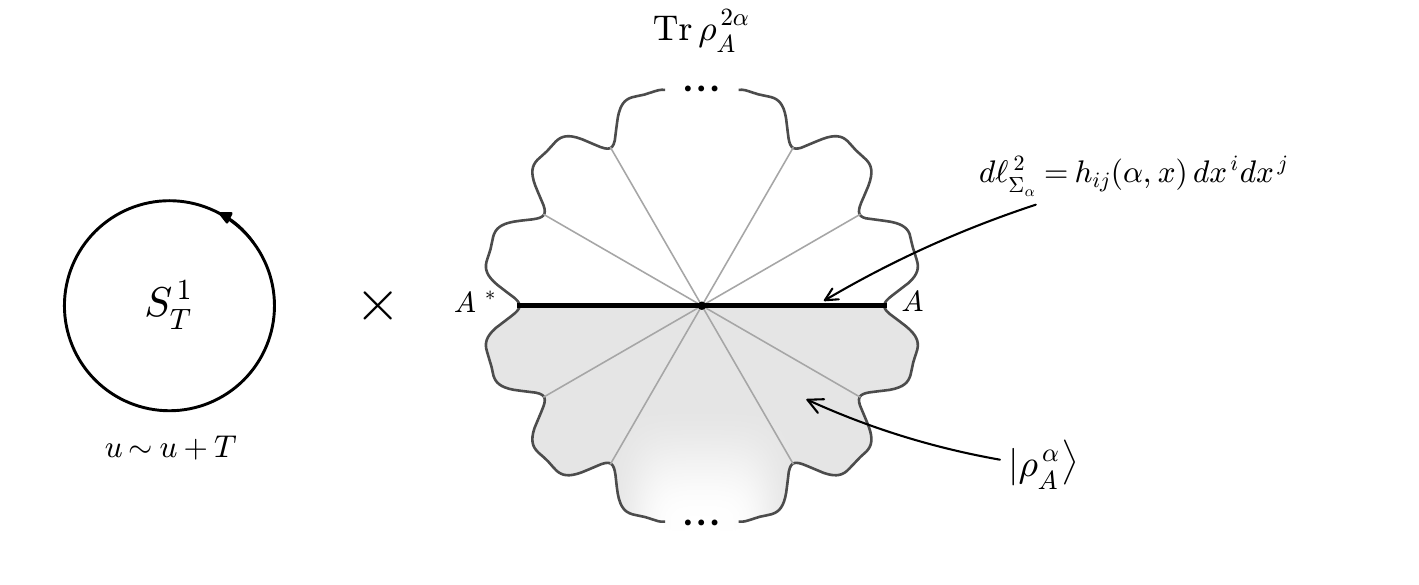}
 \caption{Replica geometry for $n=2\alpha$, with $\alpha\in\mathds{Z}_{>0}$. The spatial metric $h_{ij}(\alpha,x)$ is obtained from the time-reflection Cauchy slice of the replica-symmetric Euclidean path integral for $\operatorname{Tr}\rho_A^{2\alpha}$. The lower half prepares the canonical purification $|\rho_A^\alpha\rangle$, up to normalization.}
 \label{fig:modular_product_geometry}
 \end{figure}

For the bulk picture, we will work with classical RT geometries and treat bulk matter as quantum probes on these backgrounds, neglecting its gravitational backreaction and the displacement of the RT surface to a QES. The bulk interpretation of the purification family follows from the replica construction of \cite{DuttaFaulkner} as in figure \eqref{fig:modular_product_geometry}. 
For a Euclidean preparation with time-reflection symmetry, start with integer $\alpha$, so that $n=2\alpha$ is even. Cutting the bulk saddle for $\operatorname{Tr}\rho_A^{2\alpha}$ along its reflection-symmetric slice prepares $|\rho_A^\alpha\rangle$, up to normalization. Continuing the replica-symmetric quotient then allows $\alpha$ to vary on a chosen saddle branch; the geometry generally changes with $\alpha$. At $\alpha=1/2$, the state is the canonical purification $|\sqrt{\rho_A}\rangle$. Its bulk spatial data are obtained by gluing the entanglement wedge to its CPT conjugate across $\Sigma_A$, together with the corresponding purification of the bulk fields \cite{DuttaFaulkner}.\footnote{The classical gluing can have a discontinuity in shear, whose Lorentzian development contains an impulsive gravitational wave.}

\begin{figure}[!tbp]
\centering
\includegraphics[width=\textwidth]{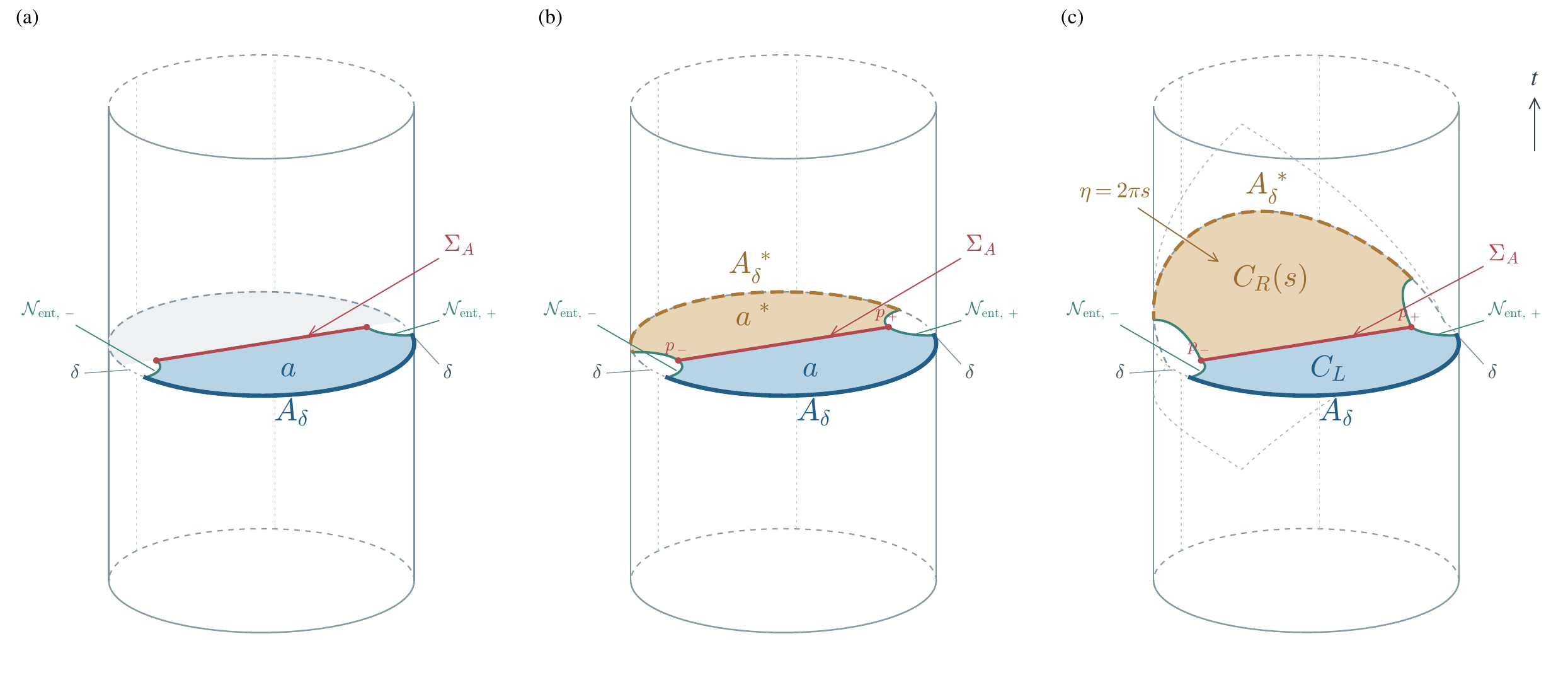}
\caption{Canonical purification and the area-generated kink, illustrated in vacuum AdS$_3$. (a) The regulated wedge $a$ on the time-symmetric $\mathbb H^2$ disk, bounded by $A_\delta$, the RT geodesic $\Sigma_A$, and tensionless branes $\mathcal{N}_{\mathrm{ent},\pm}$. (b) Gluing its CPT copy $a^*$ across $\Sigma_A$ prepares the canonical purification; the branes continue smoothly. (c) At $p_\pm=\Sigma_A\cap \mathcal{N}_{\mathrm{ent}}$, the Cauchy half-slices differ by a boost of oriented rapidity $\eta=2\pi s$ (one half held fixed). The boost is tangential to the brane and preserves $K_{ab}=0$. The intrinsic spatial metric is unchanged; full JLMS evolution also rotates the bulk fields, as in (\ref{eq:modular_purification_boosted_state}).}
\label{fig:canonical_purification_kink}
\end{figure}

We now focus on $\alpha=1/2$ and talk about the bulk dual of the corresponding one-sided modular flow. Within the semiclassical code subspace, JLMS identifies the boundary generator with the full gravitational modular generator,
\be\label{eq:modular_purification_jlms}
K_A=\mathcal{K}_{a}^{\mathrm{grav}}
=\frac{\widehat{\operatorname{Area}}(\Sigma_A)}{4G_N}
+\mathcal{K}_{bulk}^{L}+\cdots,
\ee
where $a$ is the classical entanglement wedge and $\mathcal{K}_{bulk}^{L}$ is its one-sided bulk matter modular Hamiltonian \cite{JLMS}.  Let $|\Psi_0^{\mathrm{grav}}\rangle$ be the bulk dual of $|\sqrt{\rho_A}\rangle$. Then the bulk dual of $|\Psi_{1/2,s}\rangle$ is given by
\be\label{eq:modular_purification_boosted_state}
|\Psi_s^{\mathrm{grav}}\rangle\simeq
\exp\left[-is\left(
\frac{\widehat{\operatorname{Area}}(\Sigma_A)}{4G_N}
+\mathcal{K}_{bulk}^{L}+\cdots\right)\right]
|\Psi_0^{\mathrm{grav}}\rangle.
\ee
The bulk geometry can be understood through the so-called kink transformation generated by the area term \cite{JLMS,Faulkner:2018faa,Bousso:2020yxi,Kaplan:2022orm}. Choosing a Cauchy slice through $\Sigma_A$, the area charge generates a one-sided boost in the normal plane to $\Sigma_A$, changing the relative normal frame with which the two halves of the slice are glued. In the convention of \eqref{eqn;KaZa}, the boost rapidity has magnitude $2\pi|s|$. Figure (\ref{fig:canonical_purification_kink}) illustrates the kink transformation using vacuum AdS$_3$ geometry.
The bulk field term in (\ref{eq:modular_purification_boosted_state}) supplies the accompanying one-sided bulk matter modular flow. The probe state on the doubled wedge is the canonical purification $|\sqrt{\rho_a^{\mathrm{bulk}}}\rangle$ of the original bulk reduced state \cite{DuttaFaulkner}. Assuming the original matter state has the usual vacuum short-distance behavior, its leading correlations across the RT surface have the Minkowski-vacuum entanglement structure.
This means that, near $\Sigma_A$, the leading bulk modular action is a Rindler boost. In the Killing case, the combination of the kink transformation and the bulk matter modular flow gives a smooth state with a shifted boundary time origin. JLMS argues that this local cancellation also applies without a $U(1)$ symmetry, so the full one-sided boundary modular flow is expected not to introduce a local bulk singularity \cite{JLMS}.

By definition, $|\Psi_s^{\mathrm{grav}}\rangle$ is boost-invariant, and integration over the relative boost $s$ leads to a ramp growth of the modular SFF as in \eqref{eq:modular_ramp_phase_space}, and we view this summation of semiclassical states as the generalization of the double cone wormhole to the case of general RT surfaces. In the case of the existence of U(1) isometry, the modular flow is geometric and the corresponding bulk dual is controlled by the Einstein action; this reduces to a saddle point geometry of Einstein's equations. For the more general case, the modular flow will be non-local and its bulk dual may not be controlled by Einstein's equations, but still one could imagine probing this geometry using boundary operators once the bulk modular flow can be solved. We will not do this analysis, but only study a case where the modular flow is perturbatively close to a geometric flow in section \ref{sec:two_spheres}. But before that, let's make a comment on the effect of summation of nearby states. Based on the logic of gravitational path integral, once we have a gravitational saddle, one needs to study its one-loop determinant. One nice feature of the double cone wormhole is that it has cylinder topology, and its one-loop determinant reduces to a trace over the bulk matter states. In our case, those are the states with the same classical geometry but different matter excitations, and the bulk trace is generated by the bulk modular flow:
\be\label{eq:modular_matched_generator}
K_A^L-K_A^R\simeq\mathcal{K}_{bulk},\qquad
\mathcal{K}_{bulk}
=\mathcal{K}_{bulk}^{L}-\mathcal{K}_{bulk}^{R}.
\ee
Here the superscripts $L,R$ refer to the two bulk entanglement wedges. The bulk modular evolution leaves the canonical purification invariant but acts nontrivially on nearby excited states, and we expect those to decay at late time. As in the geometric case, to obtain their late-time decay, an appropriate \(i\epsilon\) prescription will be needed. In the presence of modular \(U(1)\) symmetry, this prescription is realized by an infinitesimal deformation of the geometry near the boost fixed surface, and the resulting decay is governed by quasinormal modes. For general modular evolution, although the flow is not globally geometric, its leading behavior near the RT surface is universally described by a Rindler boost. This suggests that the same local \(i\epsilon\) deformation can be generalized to the non-geometric case. We expect the regulated modular evolution to exhibit the same decay  of nonstationary bulk fluctuations, leaving only the canonical purifications to survive at late times.

\section{\texorpdfstring{}{}Modular flow with \texorpdfstring{$U(1)$}{U(1)} symmetry}\label{sec:geometric_flow}
We begin with the case of a $U(1)$ symmetry such as thermal states and spherical subregions of CFTs. For the thermal states dominated by black holes, this is the standard double cone. So here we focus on the second case. 
To make computations explicit, we focus on the case of a 2d CFT on a circle of radius $r=1$. The Hartle-Hawking vacuum is defined by a Euclidean path-integral on the semi-infinite cylinder parametrized by: $(x,\tau) \in [0,2\pi]\times(-\infty,0]$, and we consider the subsystem $A=[0,\ell]$ where $\ell\leq \pi$, see Figure (\ref{fig:single_arc_frames}). In this case, the vacuum modular Hamiltonian takes the local form:
\be
K_A=2\pi\int_0^\ell dx\,
\frac{\cos(x-\ell/2)-\cos(\ell/2)}{\sin(\ell/2)}
\,T_{00}(0,x) - c_A
\ee
where $c_A$ is a constant shift to enforce $\text{Tr}\; e^{-K_A}=1$. To regulate the entanglement divergence at $\partial A$, we impose the boundary condition dual to a tensionless brane in AdS$_3$ as discussed in \eqref{eq:tensionless_brane_bc}.

This problem is related to the thermal state via a conformal transformation, where the regulated entanglement spectrum becomes the energy spectrum of a BCFT in the open channel description. The modular SFF thus closely follows the ordinary holographic SFF of an IR-regulated hyperbolic black hole, with a few special features due to Virasoro symmetry. We summarize the main results below so that readers familiar with the standard analysis can skip the technical details that follow.

We work with the micro-canonical modular SFF denoted by $F_A(E,T)$, where the micro-canonical window for each trace is given by a Gaussian weight:\footnote{We normalize the Gaussian window so that $\int d\mathcal{E}\,w_E(\mathcal{E})^2=\sigma_{E}$, giving the double-cone ramp $T\sigma_{E}/(2\pi)$. For the AdS$_3$ case, we impose the Gaussian weight on the primaries by dividing out the descendant contributions.}
\be
w_E(\mathcal{E})=\exp[-\pi(\mathcal{E}-E)^2/(2\sigma_{E}^2)].
\ee
At late modular time, the semiclassical result for the modular SFF, including one-loop corrections, takes the form given in (\ref{eq:moduar_SFF_total}),
\be
F_A(E,T)\sim
2\sigma_{E}^2
\exp\!\left(\sqrt{\frac{4z_\delta}{G_N}\left(E-\frac{z_\delta}{4G_N}\right)}-\sigma_{E}^2T^2/\pi\right)
+2T\int \frac{\d\mathcal{E}}{2\pi}\;\frac{w_E(\mathcal{E})^2}{|\eta(\tau_{\mathrm{dc}})|^2}.
\ee
Here $\eta$ is the Dedekind eta function, $z_{\delta}$ is determined by the UV cut-off, and $\tau_{\rm dc}$ is the modular parameter:\footnote{To make the $\eta$ function convergent, we can add a small imaginary part to $\tau_{\rm dc}$.}
\be 
\tau_{\mathrm{dc}}=-\frac{\pi T}{2z_\delta}.
\ee
The first term is the disconnected disc contribution and decays as a Gaussian in $T$. The second term comes from a double-cone wormhole contribution, which is a Lorentzian continuation of the annulus wormhole studied in \cite{Jafferis:2025yxt,Jafferis:2025jle}. In the double-cone contribution, the relative shift mode gives the linear ramp. The $\eta$ function comes from a sum over Virasoro descendants in $\text{AdS}_3/\text{CFT}_2$. Those are boundary gravitons with real normal frequencies labelled by integers $n\in \mathds{Z}_+$.  The double-cone boost acts with opposite signs on the two towers, giving the modulus squared of the eta function above.

\subsection{Conformal map to a cylinder}
\begin{figure}[htbp]
\centering
\includegraphics[width=\textwidth]{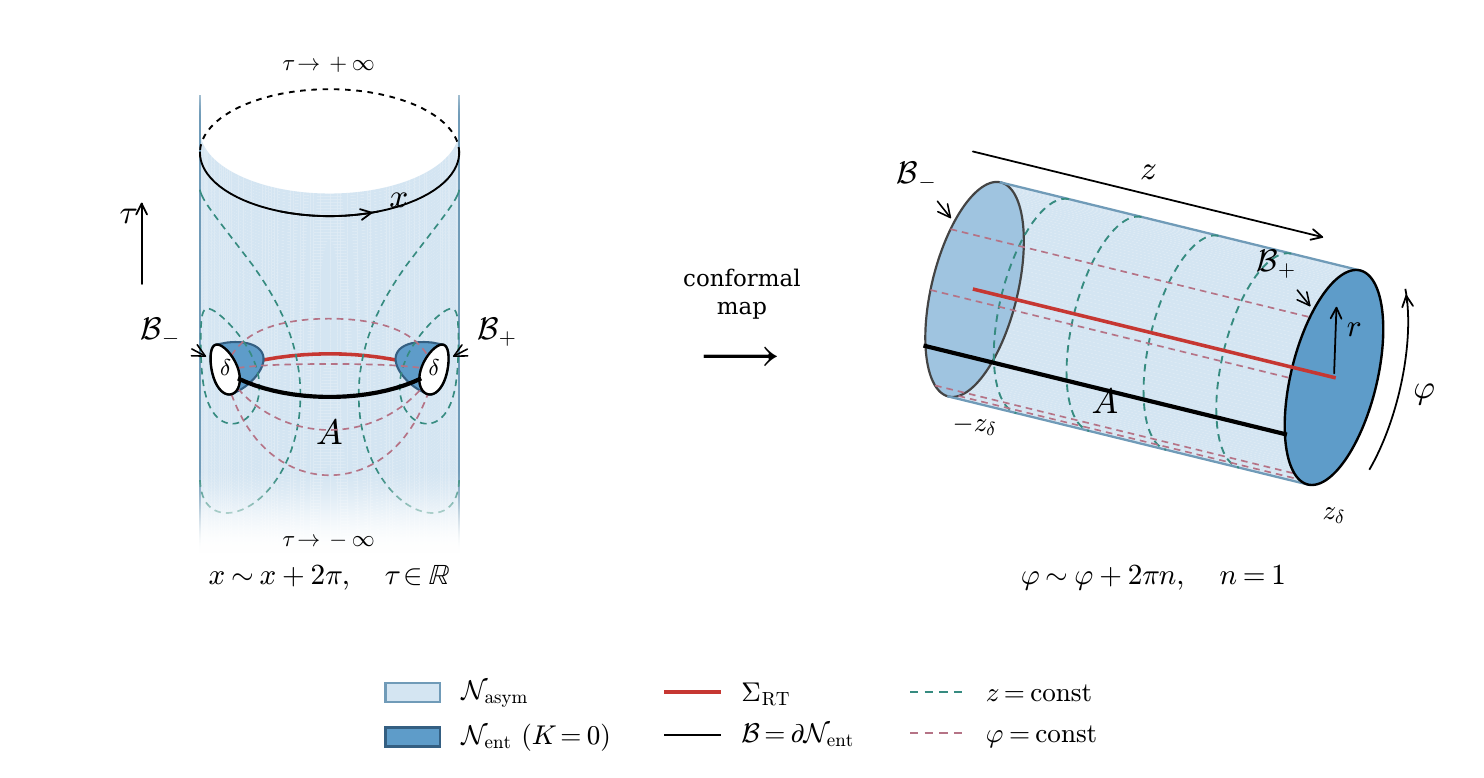}
\caption{Single-arc setup and its conformal map to the hyperbolic frame, with $n=1$ chosen for illustration.}
\label{fig:single_arc_frames}
\end{figure}

Let us first discuss the replica moment $\text{Tr} \rho^n_A$, which will be analytically continued into $\left(\text{Tr} \rho_A^{\beta\pm iT}\right)$ when we consider the disconnected contribution. It can be computed by a Euclidean path integral on the replica manifold $\Sigma^n_A$ as in \eqref{eq:modular_replica_moment}.
This path-integral admits a $U(1)$ family of conformal isometries fixing $\lbrace\partial A\rbrace$, which can be seen explicitly from the following conformal transformation into a cylinder:
\be\label{eq:conformal_map}
z+i\varphi = \ln{\left(\frac{e^{\tau+ix}-1}{e^{\tau+ix}-e^{i\ell}}\right)}, \;\;\;\d s^2= \d z^2+\d\varphi^2.
\ee
It is essentially the M\"obius transformation on the complex plane mapping $\lbrace \partial A\rbrace$ to $\lbrace 0, \infty\rbrace$, composed with the map from infinite cylinder to the complex plane. In the $z,\varphi$ variables, the conformal transformation simply shifts $\varphi$.
The subregion boundary-points $\lbrace \partial A\rbrace$ correspond to $z\to (\pm \infty)$, which requires a UV cut-off. To preserve the conformal isometry, we choose the cut-off at fixed large $z$ radius.
\be
\mathcal{B} = \mathcal{B}_+\cup \mathcal{B}_-,\;\;\;\mathcal{B}_{\pm} = \lbrace \varphi\in [0,2\pi], z=\pm z_\delta \rbrace .
\ee
In the $(x,\tau)$ coordinates, this is approximately a circle around $\lbrace \partial A\rbrace$ with radius (see Figure (\ref{fig:single_arc_frames})):
\be\label{eq:cut_off}
\delta \approx 2 \sin{\left(\frac{\ell}{2}\right)}e^{-|z|} \to z_\delta=\ln{\left[\frac{2\sin{(\ell/2)}}{\delta}\right]}.
\ee
We can define the modular parameter $\tau$ of the cylinder by a further rescaling such that the range of $z$ is $\pi$. This defines 
\be
\tau={\i \over 4 z_{\delta}}\times {\rm range~ of ~}\varphi.
\ee
In the $(z,\varphi)$ coordinates, $\Sigma^n_A$ is given by extension of the range of $\varphi$ from $[0,2\pi]$ to $[0,2\pi n]$.
In $\text{AdS}_3/\text{CFT}_2$, this gets filled by the bulk saddle of a hyperbolic black hole: 
\be\label{eq:renyi_metric}
d\ell^2 = \frac{dr^2}{r^2-n^{-2}}+(r^2-n^{-2})d\varphi^2+r^2 dz^2,\;\;\; \varphi \in [0,2\pi n], \;\;\; z\in [-z_{\delta}, z_{\delta}]
\ee
with the horizon at $r=n^{-1}$. The thermodynamical relation between the inverse temperature $n\equiv\beta$ and the normalized modular energy $E$ is given by\footnote{We distinguish the modular energy $E$ from the ADM mass $M=E-a$ of the corresponding hyperbolic black hole, measured in modular-time units.}
\be\label{eq:mass_temperature}
\beta(E)=\sqrt{\frac{a}{E-a}},\qquad a=\frac{z_\delta}{4G_N}.
\ee
The boundary $\partial\mathcal{M}$ now has two components 
$\partial \mathcal{M} = \mathcal{N}_{asym}\cup \mathcal{N}_{ent}$.
The asymptotic boundary $\mathcal{N}_{asym}$ is imposed as usual at large radius $r=r_{\text{max}}$. $\mathcal{N}_{ent}$ denotes the tensionless brane that intersects the asymptotic boundary at $\partial \mathcal{N}_{ent} = \mathcal{B}$, see Figure (\ref{fig:single_arc_frames}) for illustration.

The gravitational action including the bulk and boundary terms is given by: 
\be\label{eq:renyi_action}
I_{\Sigma^n_A}=-\frac{1}{16\pi G_N}\int_{\mathcal{M}} \sqrt{g}\left(R-2\Lambda\right)-\frac{1}{8\pi G_N}\int_{\mathcal{N}_{asym}} \sqrt{h}\;(K-1)
-\frac{1}{8\pi G_N}\int_{\mathcal{N}_{ent}}\sqrt{h}\;(K-\mathcal{T})
\ee
On this saddle point geometry (\ref{eq:renyi_metric}), the on-shell terms evaluate to:
\be 
I_{\Sigma^n_A} = -\frac{z_\delta}{4G_N n}
\ee
This gives the universal R\'enyi entropy by identifying $c=\frac{3}{2G_N}$: 
\be
S^n_A = \frac{n I_{\Sigma^1_A}-I_{\Sigma^n_A}}{1-n}  = \frac{c(n+1)}{6n}\ln{\left[\frac{2\sin{(\ell/2)}}{\delta}\right]}
\ee

Now we compute holographic contributions to the modular SFF $F_A(\beta,T)$, which is defined with two asymptotic boundary manifolds that are identified along the modular trajectories by $\varphi\sim\varphi+2\pi(\beta\pm iT)$. In this case, this is the standard analysis applied to hyperbolic black holes, which decomposes into the disconnected and connected contributions involving the disc and double-cone respectively: 
\be
F_A(\beta,T) \sim \prod_{\pm} Z_{disc}(\beta\pm iT)+ Z_{d.c}(\beta, T).
\ee

\subsection{Disconnected saddle}

\paragraph{On-shell action}
The disconnected contribution consists of the product of two disc partition functions $Z_{disc}$, each corresponding to the bulk computation for one of the traces $\text{Tr} \rho_A^{\beta\pm iT}$. Including the normalization of $\rho_A$, the action is $I_{\Sigma^n_A}-nI_{\Sigma^1_A}$ continued to $n=\beta\pm iT$, giving that:
\be\label{eq:sc_action}
I_{disc}(\beta\pm iT) = a(\beta\pm iT)-\frac{a}{\beta\pm iT}
\ee
Due to the analytic continuation of $n\to \beta \pm iT$, the geometry is complex and the horizons are moved to the complex values at $r=(\beta\pm iT)^{-1}$.  The action is UV sensitive through $z_{\delta}$ to the entanglement cut-off $\delta$. The two actions (\ref{eq:sc_action}) add up to give a total on-shell action for the disconnected contribution whose $T$-dependent part vanishes at large $T$:
\be\label{eq:disconnected_action}
I_{\text{dis}} = 2a\beta-\frac{2a\beta}{\beta^2+T^2}.
\ee
One way to get rid of the complex geometry is to consider the micro-canonical ensemble. 
With the insertion of a Gaussian micro-canonical window $w_E(\mathcal{E})=e^{-\pi(\mathcal{E}-E)^2/(2\sigma_{E}^2)}$, each of the traces evaluates to:
\be
Z_{disc}(E,\pm T)
=\int d\mathcal{E}\;w_E(\mathcal{E})
\int\frac{d\beta}{2\pi i}\,
e^{\beta\mathcal{E}}Z_{disc}(\beta\pm iT)
\simeq \sqrt{2}\sigma_{E}\nu(E)
e^{\mp iTE-\sigma_{E}^2T^2/(2\pi)},
\ee
where the $\beta$ integral finds a saddle point at $\beta(E)\mp iT$, such that the bulk geometry is the Euclidean black hole at energy $E$. 
In the semi-classical limit $a\to\infty$ with $E/a>1$ fixed, the density $\nu(E)$ is given at the leading exponential order by:
\be
\nu(E)\sim e^{2\sqrt{a(E-a)}}.
\ee
Combined together, the disconnected on-shell contribution to the micro-canonical modular SFF exhibits a Gaussian decay controlled by the width $\sigma_{E}$:
\be
Z_{disc}(E,T)Z_{disc}(E,-T)
\simeq 2\sigma_{E}^2\;
 \exp{\left(4\sqrt{a(E-a)}-\sigma_{E}^2T^2/\pi\right)}.
\ee

\paragraph{One-loop analysis}
We could also compute the one-loop corrections despite the Gaussian decay of the disconnected contribution. For the micro-canonical ensemble, since the saddle point is the original Euclidean black hole at inverse temperature $\beta(E)$, the one-loop determinant is time independent and can be calculated by the DHS formula \cite{DHS} on that background.  A slightly more interesting (but less universal) quantity is to calculate the one-loop determinant of the canonical ensemble on the complex geometry. We put that in Appendix \ref{app:gaussian_decay}.

\subsection{Double cone}

The connected saddle takes the form of a double cone, obtained from taking a quotient of the hyperbolic black hole with boost evolution $T$\cite{Saad:2018bqo,Chen:2023dc}:
\be\label{eq:dc_metric}
d\ell^2 = \frac{dr^2}{r^2-\beta(\mathcal{E})^{-2}}-4\pi^2(r^2-\beta(\mathcal{E})^{-2})dt^2+r^2 dz^2,\;\;\; t \in [0,T], \;\;\; z\in [-z_{\delta}, z_{\delta}]
\ee
Unlike the Euclidean disc, the double cone has no smoothness condition relating its Lorentzian period to the black-hole temperature, so $\mathcal{E}$ and $T$ are independent parameters.

Near the bifurcation surface, with $\rho\simeq\sqrt{2(r\beta(\mathcal{E})-1)}$, the metric is
\be\label{eq:rindler}
d\ell^2\approx d\rho^2-\frac{4\pi^2}{\beta(\mathcal{E})^2}\rho^2dt^2+\beta(\mathcal{E})^{-2}dz^2,
\qquad t\sim t+T,\quad -z_\delta\leq z\leq z_\delta.
\ee
The two exteriors correspond to opposite signs of $\rho$. The quotient has a singular fixed locus at $\rho=0$, which is avoided by a small complex deformation of the radial contour connecting the two sides. This implements the modified-boost prescription \cite{Chen:2023dc}:
\be\label{eq:modified_boost}
\widetilde{\mathcal{K}_{bulk}}=\mathcal{K}_{bulk}-i\epsilon \mathcal{H}_{bulk},
\ee
where $\mathcal{K}_{bulk}$ generates the Killing evolution and $\mathcal{H}_{bulk}$ generates local inertial-time translations near the bifurcation surface \cite{Chen:2023dc}. 

The regularized double cone has no Euclidean cap or additional boundary at the bifurcation surface. The brane contribution vanishes because $K=\mathcal{T}=0$, while the remaining Lorentzian action is
\be
S_{\mathcal{N}_{\mathrm{ent}}}=0,\qquad
S_{\mathrm{bulk}}+S_{\mathcal{N}_{\mathrm{asymp}}}=-T(E_L-E_R)=0,
\ee
since both sides share the same energy $E_L=E_R=\mathcal{E}$. The on-shell action therefore vanishes for any $\mathcal{E}$ and $T$:
\be\label{eq:dc_action}
I_{d.c}(\mathcal{E},T)=0.
\ee

The connected contribution to the modular SFF is obtained by integrating over $\mathcal{E}$. As in (\ref{eq:modular_sff_doubled_trace}), the canonical ensemble supplies the spectral weight $e^{-2\beta_{\mathrm{ext}}\mathcal{E}}$, where $\beta_{\mathrm{ext}}$ is the external inverse temperature, distinct from $\beta(\mathcal{E})$. Schematically,
\bea
Z_{d.c}(\beta_{\mathrm{ext}},T)
&\sim&\int d\mathcal{E}\;e^{-2\beta_{\mathrm{ext}}\mathcal{E}}\,\mathcal{A}(\mathcal{E},T)e^{-I_{d.c}(\mathcal{E},T)},\\
Z_{d.c}(E,T)
&\sim&\int d\mathcal{E}\;w_E(\mathcal{E})^2\,\mathcal{A}(\mathcal{E},T)e^{-I_{d.c}(\mathcal{E},T)},\label{eq:dc_ensemble_integrals}
\eea
where $\mathcal{A}(\mathcal{E},T)$ denotes the one-loop factor. Since the on-shell action vanishes for every $\mathcal{E}$, the canonical Boltzmann weight drives the energy integral toward lower energies. The micro-canonical window restricts this runaway direction to the prescribed energy range.

\paragraph{One-loop corrections}
The one-loop contribution of the double cone takes the following form: 
\be
Z_{d.c}(E,T) =\left(\text{Tr}_{T_{rel},\mathcal{E}}\right) \times \left(\text{Tr}_{\mathcal{H}_{QFT}}\; e^{-i \tilde{\mathcal{K}}_{bulk} T} \right)
\ee
 The trace over the zero-modes $\lbrace T_{rel},\mathcal{E} \rbrace$ leads to the linear ramp due to a compactification of the period of $T_{rel}$
 \be
\text{Tr}_{T_{rel},\mathcal{E}} \sim \int \frac{dT_{rel} d\mathcal{E}}{2\pi} = T\int \frac{d\mathcal{E}}{2\pi}
\ee
The integral over $\mathcal{E}$ is fixed by the micro-canonical window in defining the modular SFF. For the Gaussian kernel in (\ref{eq:dc_ensemble_integrals}), it gives:
\be
\frac{T}{2\pi}\int d\mathcal{E}\;w_E(\mathcal{E})^2
=\frac{T}{2\pi}\int d\mathcal{E}\;e^{-\pi(\mathcal{E}-E)^2/\sigma_{E}^2}
\simeq\frac{T\sigma_{E}}{2\pi},
\ee

\begin{figure}[htbp]
\centering
\includegraphics[width=0.8\textwidth]{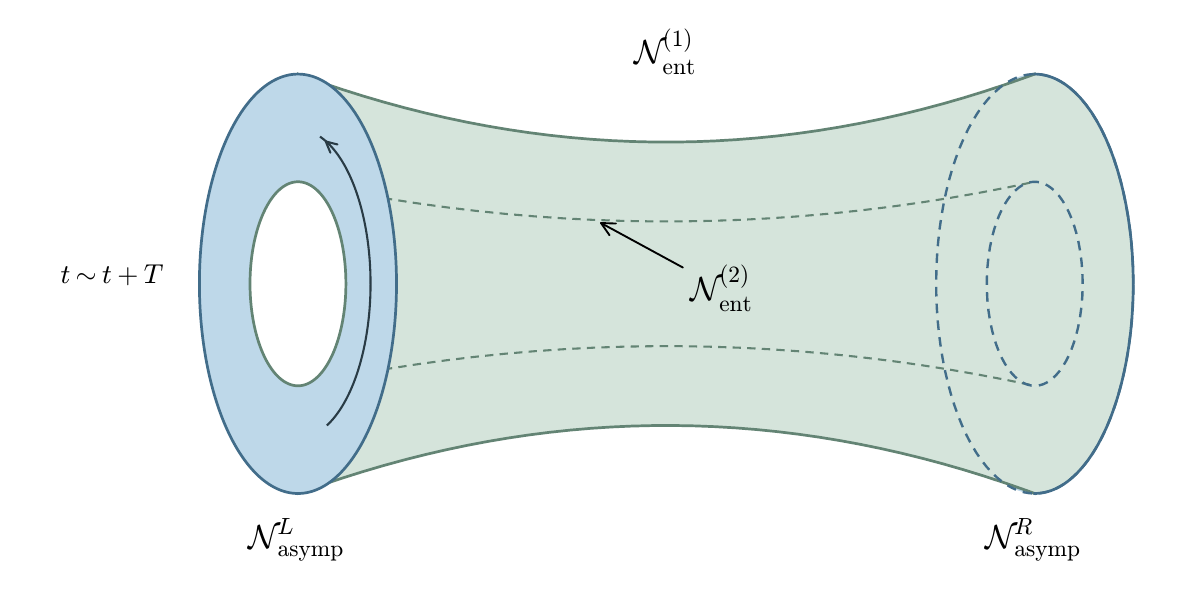}
\caption{Annulus-wormhole topology of the double-cone.}
\label{fig:annulus_wh}
\end{figure}

The bulk trace $\text{Tr}_{\mathcal{H}_{QFT}}\; e^{-i \tilde{\mathcal{K}}_{bulk} T} $ is given by a sum over quasi-normal frequencies \cite{Chen:2023dc}. In AdS$_3$ the slightly interesting contribution comes from the boundary gravitons that do not propagate into the bulk from either of the two asymptotic boundaries and lead to a set of normal frequencies $\lbrace \omega_n = (n\pi^2)/z_{\delta},n=\pm1,\pm2,...\rbrace$. These frequencies are independent of the inverse temperature, and contribute to the micro-canonical answer:
\be 
Z^{\mathrm{vir}}_{d.c}(E,T) = \frac{1}{|\eta(\tau_{\mathrm{dc}})|^2},\qquad \tau_{\mathrm{dc}}=-\frac{\pi T}{2z_\delta}.
\ee
This is an oscillatory factor that does not decay away, and is a special feature of having the Virasoro symmetry in $\text{AdS}_3/\text{CFT}_2$. To make the sum convergent, we can add a small positive imaginary part to $\tau_{\mathrm{dc}}$. This agrees from a holographic perspective with the annulus wormhole result obtained in \cite{Jafferis:2025jle}, viewed in our context as the modular Hamiltonian double-cone, see Figure (\ref{fig:annulus_wh}).

In the end, there exists an additional symmetry-twisted double-cone contribution that doubles the leading ramp coefficient, see \cite{Yan:2023rjh,Jafferis:2025jle}.

Putting both the disconnected disc contribution and the connected double-cone contribution together, we summarize the late-time evolution of the micro-canonical modular SFF as follows: 
\bea\label{eq:moduar_SFF_total}
F_A(E,T) &\sim & 2\sigma_{E}^2\;\exp{\left(\sqrt{\frac{4z_\delta}{G_N}\left(E-\frac{z_\delta}{4G_N}\right)}-\sigma_{E}^2T^2/\pi\right)}+ \frac{2T\sigma_{E}}{2\pi|\eta(\tau_{\mathrm{dc}})|^2}.
\eea

\subsection{Single-sphere in higher dimensional CFTs}
The analysis can be extended to the case of a single spherical region in the higher dimensional $\text{CFT}_{d}$ vacuum state with $d\geq 3$. The Euclidean preparation again admits the modular $U(1)$ rotation if one chooses the entanglement cutoff to preserve the $U(1)$ isometry. The disconnected and double-cone geometries can be similarly obtained from a higher-dimensional hyperbolic black hole in $\text{AdS}_{d+1}$. The one-loop corrections will involve quasi-normal frequencies that cannot be obtained in closed-form expressions.

\section{Modular SFF on two distant spheres}\label{sec:two_spheres}

In the non-$U(1)$ symmetric case, the modular flow is not geometrical away from the RT surface either because the bulk geometry does not admit an exact $U(1)$ isometry or the quantum corrections are not geometric, or both. We will only consider the case where the modular $U(1)$ symmetry is broken only by the bulk quantum state, while the classical geometry retains the corresponding isometries.\footnote{In these cases, while the classical RT surface remains unchanged as the fixed locus of the geometric isometries, the modular $U(1)$ breaking quantum corrections can displace the QES away from it \cite{Chen:2022eyi}. } This allows us to still use the double cone geometry as a background and to study its corrections from matter fields.

One option is to consider a nearby state of the previous case of a spherical region in a vacuum, or a thermal state. In such a case, we can organize the discussion as a perturbation away from the double cone geometry, where we turn on an operator source deformation $K_A=K_A^0+\lambda O_A$. Perturbatively in $G_N$, the background geometry is not changed, nor is the discussion of the ramp. The only effect of turning on this source is to change the bulk modular flow $\tilde{\mathcal{K}}_{bulk}$, which leads to a shift of the bulk vacuum state and its quasi-normal frequencies.

A more interesting case is to consider the two distant spherical regions $A=A_1\cup A_2$ in the vacuum state. In the disconnected phase, the RT surface $\Sigma^{A}_{\text{RT}}$ and entanglement wedge $a$ are unions of their single-sphere counterparts $\lbrace \Sigma^{A_{1,2}}_{\text{RT}}, a_{1,2}\rbrace$:
\be
\Sigma^{A}_{\text{RT}} = \Sigma^{A_1}_{\text{RT}}\cup \Sigma^{A_2}_{\text{RT}},\;\;\;a=a_1 \cup a_2.
\ee
Each component admits a modular $U(1)$ isometry, but the modular flow of the joint state is not geometrical due to the non-vanishing mutual information between $A_{1,2}$ coming from the matter corrections. Its modular Hamiltonian admits the following structure in the boundary CFT as the sum over two single-sphere terms and a correlational term:
\be\label{eq:correlation_form}
K_{A_1\cup A_2} = K_{A_1} \otimes \mathds{1}_{A_2}+ \mathds{1}_{A_1}\otimes K_{A_2} + \Delta K_{A_1,A_2}
\ee
where $K_{A_i}$ denotes the previously discussed single-sphere modular Hamiltonian associated with $A_i$, and $\Delta K_{A_1,A_2}$ encodes the vacuum correlation between the two spheres. In this case, if $\Delta K_{A_1,A_2}$ vanishes in the limit where $A_{1,2}$ are infinitely separated, then we have {\it two} double cones, which leads to $T^2$ growth rather than a linear in $T$ ramp for the modular SFF.
We would like to understand the transition to the linear ramp when  $\Delta K_{A_1,A_2}$ is turned on. Similar to the single-sphere case, we work with a micro-canonical version of $F_{A_1\cup A_2}(E,T)$ by fixing the modular energy $E$.

\begin{figure}[htbp]
\centering
\includegraphics[width=0.85\textwidth]{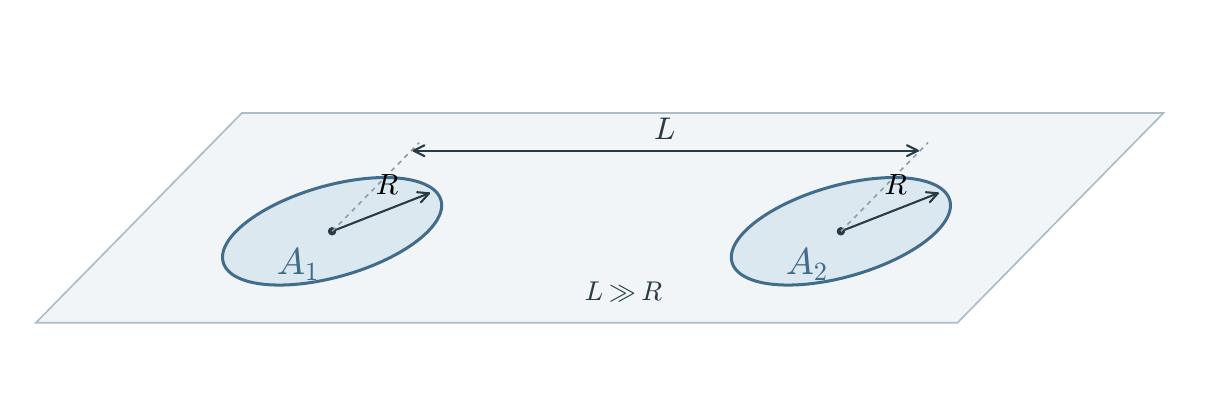}
\caption{Two-sphere sub-region $A=A_1\cup A_2$ with equal radius $R$ and a large distance $L$ between the centers.}
\label{fig:two_sphere_setup}
\end{figure}

For CFTs we can define the set-up in a conformal frame where the spheres are of equal radius $R$, separated by a distance $L$ between their centers, see Figure (\ref{fig:two_sphere_setup}). 

In the limit of large distance $L\gg R$, we can perform an OPE of $\Delta K_{A_1,A_2}$, and the leading order term is given by\footnote {Analysis for the related mutual information includes \cite{Calabrese:2010he,Cardy:2013nua,Agon:2015ftl,Chen:2022cmr,Jia:2026uuz}.}:
\be\label{eq:mod_OPE}
\Delta K_{A_1,A_2}
=-\lambda\int_{-\infty}^{\infty}ds\,f(s)O_{A_1}(s)O_{A_2}(s),\qquad
f(s)=\frac{\pi}{2\cosh^2(\pi s)},
\ee
where $\lambda=(2R/L)^{2\Delta}$ and $O_{A_i}(s)=e^{iK_{A_i}s}O_{A_i}e^{-iK_{A_i}s}$ denotes the normalized operator modular-flowed in the hyperbolic frame of $A_i$. It was derived for 2d CFTs in \cite{ChandrasekaranFaulknerLevine} by re-summing contributions from the conformal family of a primary scalar with the lowest dimension $\Delta$. We derive it in Appendix \ref{app:descendants} using an alternative OPE framework that works for general dimensions and without re-summing descendant contributions. The coupling admits a bulk EFT representation through the extrapolation dictionary, with correlators evaluated on the chosen double-cone background. It is of order $\mathcal O(1)$ in the bulk expansion, whereas $K_{A_i}$ is of order $\mathcal O(1/G_N)$. We write (\ref{eq:mod_OPE}) as $\lambda\Delta K$, and the associated Hamiltonian as $K^\lambda$. We also schematically denote $\Delta K = \mathcal{O}_1 \mathcal{O}_2$.

\subsection{Hydrodynamic action}
In the limit of $\lambda\to 0$, we have a double-cone-squared contribution, which carries the two independent ramp factors,\footnote{In the case of $\text{AdS}_3/\text{CFT}_2$, the boundary graviton sector -- one copy for each of the four boundary intervals $A^{a}_{i},a=\lbrace L,R \rbrace, i=\lbrace 1,2\rbrace$ -- remains in the late-time limit, as we have seen in the last section. We will not consider these special contributions in this section.}
\be\label{eq:early-prefactor}
Z_{dc}(E,T)=Z^{A_1}_{d.c} \cdot Z^{A_2}_{d.c}\sim T^2
\ee
This $T^2$ factor can be understood from the hydrodynamical path integral associated with the time reparametrization of each double cone \cite{Saad:2018bqo}:
\be
Z_{d.c}(E,T) \sim \int \mathcal{D} \epsilon_{1/2,L/R} \exp{\left\lbrace\i C\int_0^T \d \tau \left(\dot\epsilon_{1,L}^2+\dot\epsilon_{2,L}^2-\dot\epsilon_{1,R}^2-\dot\epsilon_{2,R}^2 \right)\right \rbrace}
\ee
where the variables $\epsilon_{1/2,L/R}$ generate time shifts at the double cone for $A_{1,2}$ respectively, and $L/R$ denotes the two sides of the double cone. They are related to the relative time shift $s_{1,2}$ and modular energy fluctuations $\mathcal{E}_{1,2}$ about $E/2$ on the two double cones as follows:
\be
s_{i}=\epsilon_{i,L}-\epsilon_{i,R},~~~\mathcal{E}_i=C(\dot \epsilon_{i,L}+\dot \epsilon_{i,R}).
\ee
Here we have imposed micro-canonical windows fixing both modular energies of $A_1$ and $A_2$ to center near $E/2$. Denoting $\beta_0\equiv\beta(E/2)$, the coefficient $C$ is given by the one-sided modular equation of state as:\footnote{For a spherical region in a $d$-dimensional CFT dual to Einstein gravity, the hyperbolic black hole gives the parametric equation of state \cite{Hung:2011nu}:
\be\label{eq:EOS}
\beta(E)=\frac{2x}{dx^2-(d-2)},\qquad
E=S_1\left(1+\frac{x^{d-2}}{2}(x^2-1)(d-1)\right),
\ee
where $x=r_h/L_{\mathrm{AdS}}$, and $S_1$ is the single-region entanglement entropy.}
\be 
C=-\frac{\beta_0}{2\beta^{\prime}(E/2)}
\ee

Now we consider turning on $\lambda \Delta K$. The relative mode $\xi_{L/R}=\epsilon_{1,L/R}-\epsilon_{2,L/R}$ couples linearly to the matter through:
\be
\lambda\int\d\tau\,\mathcal O_1(\tau+\xi/2)\mathcal O_2(\tau-\xi/2)
=\lambda\int\d\tau\,\left\lbrace\mathcal O_1(\tau)\mathcal O_2(\tau)+\xi(\tau) J(\tau)+O(\xi^2)\right\rbrace
\ee
where:
\be
J(\tau)=\frac12\left(\dot{\mathcal O}_1(\tau)\mathcal O_2(\tau)-\mathcal O_1(\tau)\dot{\mathcal O}_2(\tau)\right).
\ee
To consider their effect, we need to integrate them out on the double cone geometry. At late time, their correlators are approximated by the thermal correlators:
\be\label{eqn:corr}
S^J(\tau)=\frac12\langle\{J(\tau),J(0)\}\rangle_{\beta_0},\qquad
G^J_{LR}(\tau)=\operatorname{Tr}\!\left(\rho^{1/2}J(\tau)\rho^{1/2}J(0)\right),
\ee
with $\rho=Z^{-1}e^{-\beta_0(K_{A_1}+K_{A_2})}.$
Integrating out the matter field then leads to the following bi-local action of $\xi$:
\be
\frac{\lambda^2}{2}\int\d\tau\,\d\tau'\,\Big\{\xi_L(\tau)S^J(\tau-\tau')\xi_L(\tau')
+\xi_R(\tau)S^J(\tau-\tau')\xi_R(\tau')
-2\xi_L(\tau)G^J_{LR}(\tau-\tau')\xi_R(\tau')\Big\},
\ee
which under a derivative expansion in $\partial_\tau$ becomes:
\be\label{eq:hydro_KMS_local}
\frac{\lambda^2 G^J_{LR}(\omega=0)}2\int\d\tau\,
\left[(\xi_L-\xi_R)^2+\frac{\beta_0^2}{16}(\dot\xi_L+\dot\xi_R)^2\right]+\cdots.
\ee
In writing this we have used the KMS condition to relate $S^J$ and $G^J_{LR}$:
\be\label{eq:hydro_KMS}
S^J(\omega)=\cosh\!\left(\frac{\beta_0\omega}{2}\right)G^J_{LR}(\omega).
\ee
Here the omitted terms include derivatives of $\xi_L-\xi_R$ and higher derivatives of $\xi_L+\xi_R$.

In terms of energy variables, this induces a diffusion term for the anti-symmetric modes $(s_1-s_2)$ and $(\mathcal{E}_1-\mathcal{E}_2)$ in the hydrodynamic path integral:
\be\label{eqn:hydroac}
Z_{dc}\sim \int\mathcal{D}s_{1,2}\mathcal{D}\mathcal{E}_{1,2}
\exp\left\lbrace\int_0^T\d\tau\left[i\mathcal{E}_1\dot s_1+i\mathcal{E}_2\dot s_2
-\frac{\gamma C_+}{8}(s_1-s_2)^2-\frac{\gamma}{2C_+}(\mathcal{E}_1-\mathcal{E}_2)^2\right]\right\rbrace,
\ee
where we have defined the diffusion constant $\gamma$ and a variable $C_+$ that depends on the heat capacity:
\be\label{eq:hydro_gamma_integral}
C_+=-\frac{4}{\beta'(E/2)},\qquad
\gamma=\frac{4\lambda^2}{C_+} G^J_{LR}(\omega=0)
=\frac{4\lambda^2}{C_+}\int_{-\infty}^{\infty}\d\tau\,
\langle J_L(\tau)J_R(0)\rangle_{\beta_0}.
\ee
Based on the explicit form of (\ref{eq:mod_OPE}), the corresponding $\gamma$ is computed with:
\be
J(\tau)=-\frac12\int_{-\infty}^{\infty}ds\,f(s)
\left[\dot O_{A_1}(\tau+s)O_{A_2}(\tau+s)
-O_{A_1}(\tau+s)\dot O_{A_2}(\tau+s)\right].
\ee
This gives:
\be\label{eq:gamma_definition}
\gamma=\frac{4\lambda^2}{C_+}\int_{-\infty}^{\infty}d\tau\,
[g'_{\beta_0}(\tau)]^2,
\ee
where $g_\beta(\tau)=\langle O_L(\tau)O_R(0)\rangle_{\beta}$ is the hyperbolic thermal correlator, whose expression is
unknown for generic $\beta_0$. But at $\beta_0=1$ or equivalently setting the micro-canonical modular energy to the entanglement entropy $E=S_{A_1\cup A_2}$, one can evaluate it explicitly and find:
\be
\gamma=\frac{8\pi^{3/2}\lambda^2}
{16^{\Delta}C_+}\,
\frac{\Delta^2\Gamma(2\Delta)}{\Gamma\left(2\Delta+3/2\right)}.
\ee
We can extend (\ref{eqn:hydroac}) beyond the quadratic approximation and work out the full interaction for the antisymmetric modes at order \(\lambda^2\) and leading order in the \(\partial_\tau\) expansion, expressed in the spectral form as:
\bea\label{eq:full_V}
V &=&-i\lambda^2\int_{-\infty}^{\infty}\frac{d\omega}{2\pi}\,
G_{\beta_1}(\omega)G_{\beta_2}(-\omega)
\left[\cosh\!\left(\frac{\beta_1-\beta_2}{2}\omega\right)
-\cos\!\left(\omega(s_1-s_2)\right)\right],
\eea
where: 
\be 
\beta_i=\beta(E/2+\mathcal E_i),\;G_\beta(\omega) = \int_{-\infty}^{\infty}dt\,e^{i\omega t}g_\beta(t).
\ee
This is derived in Appendix \ref{app:full_int}. 

We will study the path integral of the action \eqref{eqn:hydroac} below. But before we move on, let's comment on the periodicity of the time shifts. In the case of $\lambda=0$, the two twist modes are compactified with periodicity $T$: $s_{1,2}\sim s_{1,2}+T$. But when we turn on the coupling $\lambda\Delta K$, this range needs to be adjusted due to the correction of modular energy $K^0 \to K^\lambda$. This then leads to an order $\lambda$ correction to the periodicity of $s_{1,2}$ relative to the period $T$ that defines $F_A(E,T)$. Such corrections will be ignored in the later discussions.

In appendix \ref{app:jt}, we report an explicit analysis in JT gravity of two coupled double cones done by AI. There, the hydrodynamical action is given by the Schwarzian action, and the two systems are coupled by a double trace deformation as in \cite{Maldacena:2018lmt}.

\subsection{Hydrodynamic path-integral}\label{sec:two_sphere_hydro}

We now study the hydrodynamic path-integral in (\ref{eqn:hydroac}). It is convenient to introduce the conjugate variables
\begin{equation}
    s_\pm=\frac{s_1\pm s_2}{2}, \qquad
    \mathcal{E}_\pm=\mathcal{E}_1\pm\mathcal{E}_2,
\end{equation}
in terms of which the effective action becomes
\begin{equation}
    \label{eq:I_eff}
    I_{\rm eff}=\int_0^T \d\tau
    \left[
    \mathcal{E}_+\dot{s}_+
    +\mathcal{E}_-\dot{s}_-
    +\frac{\i\gamma}{2}
    \left(C_+s_-^2+\frac{\mathcal{E}_-^2}{C_+}\right)
    \right].
\end{equation}

The interaction in this action was obtained by expanding around the symmetric configuration $s_-=\mathcal{E}_-=0$. The quadratic approximation is expected to be reliable at sufficiently late times $\gamma T C_+\gg 1$, when the interaction localizes the antisymmetric modes near this configuration (see \eqref{eqn:smflu}). It is not generally applicable over the entire compactified domain of the relative shifts, which is controlled by \eqref{eq:full_V}.  We will study the early- and late-time regimes separately. 
\paragraph{Early-time regime.}

We first consider modular times that are larger than the Thouless time of the individual systems, but small enough that the accumulated effect of the interaction between them is perturbative. Since the matter correlators in (\ref{eq:full_V}) decay over a finite modular time scale, the leading interaction correction to the effective action is of order $\lambda^2 T$, with an $\mathcal{O}(1)$ coefficient determined by the integrated matter correlators for $\Delta \sim \beta_{1,2}\sim \mathcal{O}(1)$.

In the weak-coupling limit, there is thus a regime $T_{thouless}\ll T\ll (\gamma C_+)^{-1}$ where we can neglect the interaction while retaining the two independent compactified shift modes.
Keeping the micro-canonical windows imposed at $\lambda=0$, the path-integral reduces to
\begin{align}\label{eq:early_Zdc}
    Z_{dc}(E,T)
    \simeq
    \frac{T^2}{(2\pi)^2}
    \int \d\mathcal{E}_1\d\mathcal{E}_2\,
    w_E^2 =\frac{T^2\sigma_+\sigma_-}{8\pi^2}.
\end{align}
where we introduced two Gaussian filters, one for each of the energy.
This recovers the product of two independent double-cone contributions, giving rise to the $T^2$ growth.

\paragraph{Late-time regime.}

At sufficiently late times, the interaction suppresses configurations away from the symmetric point $s_-=\mathcal{E}_-=0$. We can then evaluate the path-integral by expanding around this configuration using the quadratic action (\ref{eq:I_eff}). The symmetric sector remains flat and its phase-space integral gives
\begin{equation}
    Z_+(E,T)=\frac{T\sigma_+}{2\pi}.
\end{equation}
For the antisymmetric sector, we decompose the collective coordinates into their zero modes and periodic fluctuations:
\begin{equation}
    s_-(\tau)=\bar{s}_-+\delta s_-(\tau),\qquad
    \mathcal{E}_-(\tau)=
    \bar{\mathcal{E}}_-+\delta\mathcal{E}_-(\tau).
\end{equation}
The interaction localizes the zero modes within range
\begin{equation}\label{eqn:smflu}
    \Delta\bar{s}_-\sim\frac{1}{\sqrt{\gamma TC_+}},
    \qquad
    \Delta\bar{\mathcal{E}}_-
    \sim\sqrt{\frac{C_+}{\gamma T}}.
\end{equation}
Therefore for $T\gg {1\over \gamma C_+}$, we can extend the integration domain of the antisymmetric zero modes to the entire real plane and neglect the variation of the micro-canonical window within the localized region. The zero-mode integral then gives:
\begin{equation}
    Z^{0}_{-}
    =
    \int\frac{\d\bar{s}_-\d\bar{\mathcal{E}}_-}{2\pi}
    e^{-\frac{\gamma T}{2}
    \left(C_+\bar{s}_-^2+
    \bar{\mathcal{E}}_-^2/C_+\right)}
    =\frac{1}{\gamma T}.
\end{equation}
Expanding the periodic fluctuations in Fourier modes,
the Gaussian integral over the non-zero modes,
normalized by the corresponding integral at
$\lambda=0$, gives
\begin{equation}
    Z^{\mathrm{fluc}}_{-}
    =
    \prod_{n=1}^\infty
    \left[
    1+\left(\frac{\gamma T}{2\pi n}\right)^2
    \right]^{-1}
    =
    \frac{\gamma T/2}{\sinh(\gamma T/2)}.
\end{equation}
Combining the zero-mode and non-zero-mode
contributions, we obtain
\begin{equation}
    Z_-=\frac{1}{2\sinh(\gamma T/2)}.
\end{equation}
This result can also be understood by viewing the
antisymmetric phase-space path-integral as the
thermal partition function of a harmonic oscillator
with Hamiltonian
\begin{equation}
    H_-=\frac{\gamma}{2}
    \left(C_+\hat{s}_-^2+
    \frac{\hat{\mathcal{E}}_-^2}{C_+}\right),
\end{equation}
whose spectrum is given by
$\gamma(n+1/2)$.

There is, however, a subtlety in evaluating the
non-zero-mode determinant. The hydrodynamic action
is valid only at low frequencies, whereas the
infinite product above involves arbitrarily high
frequencies. We account for the contribution from
the high-frequency modes by introducing a
counter-term $I_{\rm ct}$. On a stationary
background, the leading correction is extensive
in $T$ and takes the form of a shift of the
effective vacuum energy. Based on unitarity, the vacuum energy should exactly vanish. This means there must exist the following counter-term:
\begin{equation}
    \label{eq:ct_proposal}
    I_{\rm ct}=-\frac{\gamma T}{2}.
\end{equation}
The renormalized contribution from the antisymmetric sector is then
\begin{equation}
    Z_-^{\rm ren}
    =e^{-I_{\rm ct}}Z_-
    =\frac{1}{1-e^{-\gamma T}}.
\end{equation}
For $\gamma T\gg1$, this contribution approaches
unity, and we recover a single linear ramp:
\begin{equation}
    \label{eq:matching}
    Z_{dc}^{\rm ren}(E,T)
    \simeq\frac{T\sigma_+}{2\pi}.
\end{equation}

\section{Discussion and outlook}\label{sec:discussion}
In this paper, we have studied the ramp of the modular SFF in AdS/CFT to understand better the general connection between RT surfaces and random matrix statistics. In section \ref{sec:modular_SFF}, we proposed a bulk description in terms of a sum over boost-invariant semiclassical states with a periodic identification of the boost time. For regions with modular $U(1)$ symmetry, the computation reduces to the standard SFF of hyperbolic black holes, providing a direct realization of the double-cone contribution. To study departures from this symmetry, we focused on those induced by bulk quantum correlations and developed a perturbative treatment, using two distant spherical sub-regions as the principal example. We identified the transition from the $T^2$ behavior of two decoupled modular systems to the single RMT ramp proportional to $T$ due to these correlations. In the absence of correlations, each sphere supports an independent set of collective zero-modes and slow fluctuations. Using perturbation theory formulated for general weakly coupled modular Hamiltonians, we showed that correlations generate a diffusive coupling that lifts the antisymmetric zero-modes and suppresses their fluctuations at large $T$, leaving only the symmetric sector to support the surviving ramp. This analysis identifies a mechanism through which bulk quantum correlations reorganize entanglement level statistics even when the classical entanglement wedge remains disconnected. It also provides perturbative support for the proposal in section \ref{sec:modular_SFF}: although the correlated boundary modular flow is nongeometric, the bulk description retains a sum over boost-invariant states with a periodic identification of the common boost time.

Our study of the two sphere case could be extended to more general situations when we study the spectral form factor of weakly coupled chaotic systems, such as a weakly coupled chaotic chain. At late time and finite volume, it should reduce to the hydrodynamical action with the inclusion of the spatial diffusion, with $\gamma(k)$ now depending on the spatial momentum $k$.  But recall that the quadratic expansion is only valid if the combination of $\gamma T C_+$ is large. This could break down due to two reasons: for a finite volume system,  it can break down at early time, or if the volume is infinite, then this might break down for the low momentum modes whose relaxation rate $\gamma(k)$ becomes small. In both cases, the compactification of the twist modes will be important, and it would be nice to understand its consequence.

In principle, we should go beyond the two-sphere case and investigate modular flow for other states and sub-regions without modular $U(1)$ symmetry. It would be interesting to understand this case better. It seems also interesting to consider the modular SFF for the Hawking radiation after the Page transition. The construction in section \ref{sec:modular_SFF} will lead to a double cone in a closed universe, in the spirit of the de Sitter double cone \cite{Yang:2025lme}. A concrete starting point is the PSSY model \cite{Penington:2019kki}, where one can study the modular SFF of the radiation by considering gluing the pinwheel geometries using the double trumpet.

\acknowledgments
We thank Guancheng Lu for initial collaborations. We thank Frederic Jia, Arvin Shahbazi-Moghaddam and Douglas Stanford for discussions. This work is supported by National Science Foundation of China (NSFC) grant No. 12447108.
ZY is also supported by NSFC Grant No.12342501, and by Tsinghua University Dushi Program (No.20251080043 and No.20261080110).

\appendix

\section{Full interaction $V(s_\pm,\mathcal{E}_\pm)$ at $\lambda^2$ order}\label{app:full_int}
To understand the full interaction between the collective modes $s_\pm$ and $\mathcal{E}_\pm$, we need to work out the matter path-integral: 
\be\label{eq:matter_PI}
e^{iI_{\mathrm{eff}}}\propto \int\mathcal{D}\phi\; e^{\i \int^T_0 \d\tau \left(\dot{\phi} \cdot \Pi_\phi-K_{d.c}^{\lambda}\right) }
\ee
at a fixed background $\vec{q}(\tau)=\lbrace s_{\pm}(\tau),\mathcal{E}_{\pm}(\tau) \rbrace$, where: 
\be
K^\lambda_{d.c} = K^{\lambda,L}_{A_1\cup A_2} - K^{\lambda,R}_{A_1\cup A_2}+\mathcal{O}(i\epsilon)
\ee
should be understood via its representation in the bulk EFT. We can evaluate (\ref{eq:matter_PI}) in the Hamiltonian formalism by defining the bulk EFT Fock space: 
\be\label{eq:Fock}
\mathcal{H}_{\mathrm{QFT}}(\vec{q})
=\operatorname{span}\left\{\left(\prod_j\hat\Phi_j^\dagger\right)
|\Omega^\lambda_{\vec{q}}\rangle\right\}.
\ee
Here $\hat\Phi_j^\dagger$ creates a bulk-field excitation on the bulk EFT vacuum $|\Omega^\lambda_{\vec q}\rangle$ of $K^\lambda_{d.c}$ on the fixed spatial geometry labeled by $\vec{q}=\lbrace (s_1\pm s_2)/2, \mathcal{E}_1\pm \mathcal{E}_2 \rbrace$. In more detail, the spatial geometry consists of a pair of Cauchy slices, one in each two-sided hyperbolic black hole with modular energy $E/2+\mathcal{E}_i$, and relative left-right time shift $s_i$.

We can therefore rewrite (\ref{eq:matter_PI}) by discretizing the time-circle as: 
\be\label{eq:Delta_I_def}
\lim_{N\to\infty}
\operatorname{Tr}_{\mathcal{H}_{\mathrm{QFT}}(\vec{q})}\left[
\prod_{i=0}^{N-1}
\left(e^{-iK^\lambda_{d.c}\delta\tau}\right)^{\mathcal{H}_{\mathrm{QFT}}(\vec{q}_{i+1})}_{\mathcal{H}_{\mathrm{QFT}}(\vec{q}_i)}\right],
\ee
where $\delta\tau=T/N$, $\tau_i=i\delta\tau$, $\vec{q}_i=\vec{q}(\tau_i)$, and $\vec{q}_N=\vec{q}_0$.  In writing (\ref{eq:Delta_I_def}) we have defined the ``transitional projection'' operator by:
\be\label{eq:transitional_projection}
\left(e^{-iK^\lambda_{d.c}\delta\tau}\right)^{\mathcal{H}_{\mathrm{QFT}}(\vec{q}_{i+1})}_{\mathcal{H}_{\mathrm{QFT}}(\vec{q}_i)}
=\sum_{\vec{\phi}_i,\vec{\phi}_{i+1}}
\langle\vec{\phi}_{i+1}|e^{-iK^\lambda_{d.c}\delta\tau}|\vec{\phi}_i\rangle
\cdot |\vec{\phi}_{i+1}\rangle\langle\vec{\phi}_i|,
\ee
where $\lbrace \vec{\phi}_i\rbrace$ and $\lbrace\vec{\phi}_{i+1}\rbrace$ denote collectively the complete orthonormal bases of the neighboring Fock spaces $\mathcal{H}_{\mathrm{QFT}}(\vec{q}_i)$ and $\mathcal{H}_{\mathrm{QFT}}(\vec{q}_{i+1})$.

The dissipative nature of $K^\lambda_{d.c}$ is characterized by the Thouless time. In the limit of large $T$ and slowly varying $\vec{q}(\tau)$, we can approximate (\ref{eq:Delta_I_def}) at a $\delta \tau$ satisfying: 
\be 
T_{thouless}\ll\delta\tau \ll \tau_{\vec{q}}
\ee
where $\tau_{\vec{q}}$ is the characteristic timescale of variation for $\vec{q}(\tau)$, so that $\vec{q}(\tau)$ is effectively constant during an interval of $\delta \tau$. By doing this, the transitional projection is dominated by the instantaneous ground states:
\be 
\left(e^{-iK^\lambda_{d.c}\delta\tau}\right)^{\mathcal{H}_{\mathrm{QFT}}(\vec{q}_{i+1})}_{\mathcal{H}_{\mathrm{QFT}}(\vec{q}_i)}
\approx e^{-i E^\lambda_{\Omega}(\vec{q}_i)\delta\tau}
\cdot |\Omega^\lambda_{\vec{q}_{i+1}}\rangle\langle \Omega^\lambda_{\vec{q}_i}|.
\ee
The excited contributions dissipate away exponentially -- we neglect special cases that allow normal frequencies. This is the adiabatic approximation, and gives the leading order in small $\partial_\tau$ contribution of the matter path-integral (\ref{eq:matter_PI}) -- by tracking the instantaneous  vacuum eigenvalue of $K^{\lambda}_{d.c}$ in the Fock space $\mathcal{H}_{\text{QFT}}(\vec{q}(\tau))$ \footnote{We emphasize that the exact periodicity of the shifts $s_{1,2}$ on the finite time circle is encoded in the excited-state contributions omitted by the large-$T$ vacuum approximation.}:
\be\label{eq:two_sphere_potential_def}
\int \mathcal{D}\phi\;e^{i\int^T_0 d\tau \left(\dot{\phi}\cdot \Pi_{\phi}-K^{\lambda}_{d.c}\right)} \approx e^{-i\int^T_0 d\tau E^{\lambda}_{\Omega}(\vec{q}(\tau))}
\ee
The vacuum energy as a function of the collective modes $\vec{q}$ then effectuates a coupling $V(\vec{q}) = E^\lambda_{\Omega}(\vec{q})$ between them. 

We can perform a perturbation theory in $\lambda$ of the vacuum energy: 
\be 
E^{\lambda}_{\Omega}(\vec{q}) = E^{(0)}_{\Omega}(\vec{q})+\lambda E^{(1)}_{\Omega}(\vec{q})+\lambda^2 E^{(2)}_{\Omega}(\vec{q})+...
\ee
for the perturbed double-cone modular Hamiltonian:
\be
K^{\lambda}_{d.c} = K^{0}_{d.c}+\lambda \Delta K_{d.c},\;\;\;\Delta K_{d.c} = \Delta K^L-\Delta K^R
\ee
The perturbation series begins with the product form of the ground state at $\lambda=0$:
\be\label{eq:prod_vac}
|\Omega^0_{\vec q}\rangle
=e^{-iK^L_{A_1}s_1}|\Omega\rangle_{E/2+\mathcal E_1}
\otimes e^{-iK^L_{A_2}s_2}|\Omega\rangle_{E/2+\mathcal E_2},
\ee
and the corresponding energy $E^{(0)}_\Omega(\vec q)=0$. Here $|\Omega\rangle_{E/2+\mathcal E_i}$ is the Hartle-Hawking state on the corresponding hyperbolic black hole, dual to the CFT micro-canonical thermofield double of $K_{A_i}$ at modular energy $E/2+\mathcal E_i$. 

The first order correction $E^{(1)}_\Omega(\vec{q})=\langle \Delta K^L-\Delta K^R\rangle_{\Omega^{0}_{\vec{q}}}$ vanishes identically, since one-sided quantities do not depend on $s_{1,2}$ when evaluated in (\ref{eq:prod_vac}). So the leading order correction is $\lambda^2 E^{(2)}_\Omega(\vec{q})$ with:
\bea\label{eq:vacuum_second_order_integral}
E^{(2)}_\Omega(\vec q)
&=&-i\int_0^\infty\d\tau\,
\left\langle\Delta K_{d.c}(\tau)\Delta K_{d.c}(0)\right\rangle_{\Omega^0_{\vec q}}\nonumber\\
&=&-i\int_{-\infty}^\infty\d\tau\,\left[
\left\langle\Delta K^L(\tau)\Delta K^L(0)\right\rangle_{\Omega^0_{\vec q}}
-\left\langle\Delta K^L(\tau)\Delta K^R(0)\right\rangle_{\Omega^0_{\vec q}}\right]
\eea
where we have defined $\Delta K^{L,R}(\tau)=e^{iK^0_{d.c}\tau}\Delta K^{L,R}e^{-iK^0_{d.c}\tau}$. In our case, $\Delta K^{L,R}$ are supported away from the hyperbolic horizons -- they occur near the asymptotic boundaries. So the action of  $K^0_{d.c}$ is not affected by the $i\epsilon$ part, and coincides with the local modular flows on $A_i$. 

Substituting (\ref{eq:mod_OPE}) into (\ref{eq:vacuum_second_order_integral}) and denoting $\beta_i=\beta(E/2+\mathcal E_i)$, we obtain that:
\be\label{eq:two_sphere_correlator_integral}
E^{(2)}_\Omega(\vec q)
=i\int_{-\infty}^{\infty}d\tau\,
\left[g_{\beta_1}^{LR}(\tau+s_-)g_{\beta_2}^{LR}(\tau-s_-)
-g_{\beta_1}^{LL}(\tau)g_{\beta_2}^{LL}(\tau)\right],
\ee
where $g_{\beta_i}^{ab}(t)=\langle O_{A_i}^a(t)O_{A_i}^b(0)\rangle_{\beta_i}$ are hyperbolic thermal correlators related by:
\be\label{eq:two_sphere_thermal_correlators}
g_\beta^{LL}(t)=g_\beta^{LR}\!\left(t+\frac{i\beta}{2}-i0^+\right).
\ee
Eq. (\ref{eq:two_sphere_correlator_integral}) can then be Fourier-transformed into the spectral representation:
\be\label{eq:two_sphere_energy_integral}
E^{(2)}_\Omega(\vec q)
=-i\int_{-\infty}^{\infty}\frac{d\omega}{2\pi}\,
G_{\beta_1}(\omega)G_{\beta_2}(-\omega)
\left(\cosh(\beta_-\omega)-\cos(2\omega s_-)\right),
\ee
where $\beta_-=(\beta_1-\beta_2)/2$, and $G_\beta(\omega)$ is the Fourier transform:
\be\label{eq:fourier_mode_two_sphere}
G_\beta(\omega)=\int_{-\infty}^{\infty}dt\,e^{i\omega t}g_\beta^{LR}(t)
\ee
At $\beta=1$ it is given explicitly by:
\be
G_1(\omega)=\frac{1}{2\pi\Gamma(2\Delta)}
\left|\Gamma\!\left(\Delta+\frac{i\omega}{2\pi}\right)\right|^2.
\ee
The coupling $V(\vec{q})= \lambda^2 E^{(2)}_\Omega(\vec{q})+\mathcal{O}(\lambda^3)$ through (\ref{eq:two_sphere_energy_integral}) therefore specifies the full interaction among $(s_\pm,\mathcal{E}_\pm)$ at order $\lambda^2$. It is purely dissipative: 
\be
\text{Re}\; E^{(2)}_\Omega(\vec{q})=0,\;\;\text{Im}\; E^{(2)}_\Omega(\vec{q})\leq 0
\ee
whose imaginary part admits a global maximum at the symmetric point $s_-=\mathcal{E}_-=0$. The quadratic expansion near this gives the diffusive term in (\ref{eqn:hydroac}):
\be\label{eq:quad_approx}
V(\vec q)\simeq-\frac{i\gamma}{2}
\left(C_+s_-^2+\frac{\mathcal E_-^2}{C_+}\right),
\ee
with the coefficients:
{
\be\label{eq:appendix_gamma_kernel}
C_+=-\frac{4}{\beta'(E/2)},\qquad
\gamma=\frac{4\lambda^2}{C_+}\int_{-\infty}^{\infty}\frac{d\omega}{2\pi}\,
\omega^2G_{\beta_0}(\omega)^2,
\ee
in agreement with (\ref{eq:gamma_definition}).
}

\begin{figure}[htbp]
\centering
\includegraphics[width=0.72\textwidth]{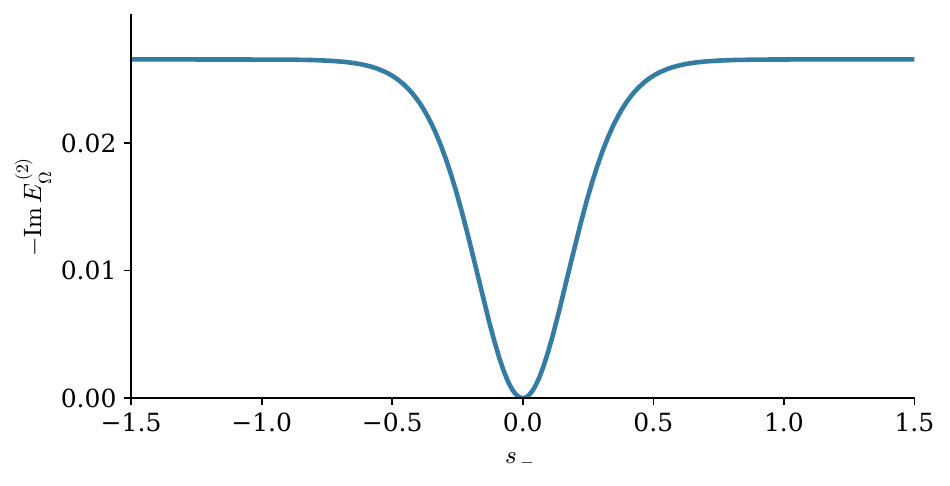}
\caption{{The coupling $-\operatorname{Im}E^{(2)}_\Omega$ at $\Delta=1$ and $\beta_1=\beta_2=1$.}}
\label{fig:two_sphere_potential}
\end{figure}

We can estimate the global landscape of $E^{(2)}_\Omega(\vec{q})$. At fixed $\beta_{1,2}$, it depends on large $|s_-|$ through the $\omega$-pole of the integrand that is closest to the real-axis, and flattens exponentially to the asymptotic value: 
\be
\lim_{s_-\to \pm\infty}E^{(2)}_\Omega(\vec{q})-E^{(2)}_\Omega(\vec{q}) \sim e^{-2\kappa |s_-|}
\ee
where $\kappa$ is the smallest imaginary part of $\omega$-poles. See Figure (\ref{fig:two_sphere_potential}) for its plot as a function of $s_-$ at the special point $\beta_1=\beta_2=1$. 

\section{A toy model in JT gravity}\label{app:JT}\label{app:jt}

In this appendix, we study coupled JT gravity theories.\footnote{The detailed calculations in this section are done by AI.} We consider two identical JT systems $H_1$ and $H_2$ coupled by a double-trace operator described by the total Hamiltonian: 
\be\label{eq:jt_model}
H=H_1+H_2+\lambda O_1O_2,
\qquad m^2=\Delta(\Delta-1),
\ee
where $O_{1,2}$ are boundary operators dual to a real bulk scalar of mass $m$. We study the micro-canonical version of the SFF associated with $H$ at energy $E$, particularly the contribution from a double-cone-squared geometry $Z_{dc}(E,T)$. 

In JT gravity, the bulk degrees of freedom are all fixed, leaving only the boundary reparameterization and geometric moduli as the gravitational variables. For the double-cone-squared geometry, these include the embedding functions $X_{i,a}(\tau)$ with $i=1,2$ and $a=L,R$, and the throat sizes $b_i$ enforcing that $X_{i,a}(\tau+T) = X_{i,a}(\tau)+b_i$. The contribution $Z_{dc}(E,T)$ is computed by the path-integral: 
\be 
Z_{dc}(E,T) = \int db_i \mu(b_i)\;{\hat{w}_{E}^{\,2}} \int \mathcal{D} X_{i,a}(\tau) \int \mathcal{D}\Phi_i\; e^{-I(X,\Phi)} 
\ee 
with an explicit action given by:
\be\label{eq:jt_recap_action}
I(X,\Phi)=-C\sum_{i,a}\int d\tau\,\{e^{X_{i,a}(\tau)},\tau\}
+\sum_{i} I_{\Phi_i}+\lambda\sum_a\int d\tau\,O_{1,a}O_{2,a},
\ee
where $C$ is the Schwarzian coupling, and $I_\Phi$ denotes the free bulk scalar action. {The combined Gaussian weight $\hat{w}_{E}^{\,2}$, with $\hat{w}_{E}=\exp[-\pi(\bar{\mathcal{E}}_+-E)^2/(2\sigma_+^2)]$, is also inserted to implement the micro-canonical kernel of the main text.} The double-trace $O_1 O_2$ acts as a source term that couples the bulk scalars $\Phi_{1,2}$. To proceed,  we first integrate out the scalars $\Phi_{1,2}$, and then expand the result to the second order $\mathcal{O}(\lambda^2)$. We obtain that: 
\bea\label{eq:jt_effective_action}
I_{\mathrm{eff}}[X]&=&-\log Z_\Phi[X]+iC\sum_i\int_0^Td\tau\,
\left(\{e^{X_{i,L}(\tau)},\tau\}-\{e^{X_{i,R}(\tau)},\tau\}\right)\nonumber\\
&&+\frac{\lambda^2}{2}\sum_{a,b}(2\delta_{ab}-1)\int_0^Td\tau\,d\tau'\;
G_1^{ab}[X_1](\tau,\tau')G_2^{ab}[X_2](\tau,\tau')+\mathcal O(\lambda^4).
\eea
For Lorentzian embeddings $X_{i,a}(\tau)$, the propagators are given by:
\be\label{eq:jt_boundary_propagators}
G_i^{ab}[X_i](\tau,\tau')=\sum_{m\in\mathbb Z}
\left[\frac{\dot X_{i,a}(\tau)\dot X_{i,b}(\tau')}
{4\cosh^2\!\left(\frac{X_{i,a}(\tau)-X_{i,b}(\tau')+mb_i+i\pi\delta_{ab}}{2}\right)}\right]^\Delta.
\ee
Here dots and primes denote derivatives with respect to $\tau$. The scalar determinant $Z_\Phi[X]$ is evaluated at $\lambda=0$ and is given by \cite{Chen:2023dc}: 
\be
Z_\Phi[X] = \prod^2_{i=1} \prod^\infty_{k=0} \frac{1}{1-e^{-(\Delta+k)b_i}}
\ee
For $b_i\propto T\to \infty$, the determinant $Z_\Phi[X]$ approaches unity and will be ignored. In this limit, we only keep the $m=0$ contribution for $G^{ab}_i$ in (\ref{eq:jt_boundary_propagators}). 

To treat the remaining path-integral over $X_{i,a}(\tau)$, we perform the Gaussian approximation by expanding $X_{i,a}(\tau)$ near classical saddle-point solutions:
\be\label{eq:jt_saddle_expansion}
X_{i,a}(\tau)=\frac{2\pi}{\beta_i}\left[\tau+\epsilon_{i,a}(\tau)\right],\;\;\epsilon_{i,a}(\tau+T)= \epsilon_{i,a}(\tau)
\ee
where $\beta_i = 2\pi T/b_i$ are bulk inverse temperatures, and $\epsilon_{i,a}(\tau)$ are periodic fluctuations. Expanding (\ref{eq:jt_effective_action}) to quadratic order and integrating the Schwarzian term by parts gives:
\be\label{eq:jt_time_action}
\begin{aligned}
I_{\mathrm{eff}}^{(2)}={}&-\frac{iC}{2}\sum_i\int_0^T d\tau\,
\left[\ddot\epsilon_{i,L}(\tau)^2+\frac{4\pi^2}{\beta_i^2}\dot\epsilon_{i,L}(\tau)^2-(L\leftrightarrow R)\right]\\
&+\frac{\lambda^2}{4}\sum_{a,b}(2\delta_{ab}-1)\int_0^T d\tau\int^T_0 d\tau'\;
 g_1^{ab}(\tau-\tau')g_2^{ab}(\tau-\tau')\\
&\times\Bigg\{\bigg[\sum_i\Big(\Delta[\dot\epsilon_{i,a}(\tau)+\dot\epsilon_{i,b}(\tau')]+[\epsilon_{i,a}(\tau)-\epsilon_{i,b}(\tau')]\partial_\tau\log g_i^{ab}(\tau-\tau')\Big)\bigg]^2\\
&\quad+\sum_i\Big([\epsilon_{i,a}(\tau)-\epsilon_{i,b}(\tau')]^2\partial_\tau^2\log g_i^{ab}(\tau-\tau')-\Delta[\dot\epsilon_{i,a}(\tau)^2+\dot\epsilon_{i,b}(\tau')^2]\Big)\Bigg\}.
\end{aligned}
\ee
Here the background propagator $g^{ab}_i$ is given by:
\[
g_i^{ab}(\tau-\tau')=
\left[\frac{\pi^2}{\beta_i^2\cosh^2\!\left(\frac{\pi(\tau-\tau')}{\beta_i}+\frac{i\pi\delta_{ab}}{2}\right)}\right]^\Delta.
\]
Similar to the main text, we focus on expansions near the equal-temperature solutions:
\be 
\beta_- = \frac{1}{2}(\beta_1-\beta_2) \ll \beta_+ =\frac{1}{2}(\beta_1+\beta_2)
\ee
In this limit, the sum and difference of energies $\bar{\mathcal{E}}_\pm = \bar{\mathcal{E}}_1\pm \bar{\mathcal{E}}_2$ take the form:
\be\label{eq:jt_equation_of_state}
\bar{\mathcal{E}}_+ = \frac{4\pi^2 C}{\beta_+^2},\;\;\bar{\mathcal{E}}_- = -\frac{2\bar{\mathcal{E}}_+}{\beta_+} \beta_- 
\ee
The micro-canonical {Gaussian filter} ${\hat{w}_{E}^{\,2}}$ fixes that: 
\be
\int db_i \mu(b_i){\hat{w}_{E}^{\,2}} (...) \to {\sigma_+} \int d\bar{\mathcal{E}}_- \cdot (...)_{\bar{\mathcal{E}}_+ \to E}
\ee
The fluctuations can be converted into conjugate pairs $(\delta s_\pm,\delta \mathcal{E}_\pm)$:
\bea\label{eq:jt_common_relative}\label{eq:jt_energy_variables}\label{eq:jt_energy_fluctuations}
\delta s_\pm(\tau)&=&\frac12\left[\epsilon_{1,L}(\tau)-\epsilon_{1,R}(\tau)
\pm \epsilon_{2,L}(\tau)\mp \epsilon_{2,R}(\tau)\right],\nonumber\\
\delta \mathcal E_\pm(\tau)&=& \delta \bar{\mathcal{E}}_{\pm}+
\frac{E}{2}
\sum_{a=L,R}\big[\epsilon'_{1,a}(\tau)\pm\epsilon'_{2,a}(\tau)\big].
\eea
To evaluate the Gaussian integral, we collect the Fourier components of the fluctuations into:
\be
\vec X_\pm(n)=\lbrace\delta\tilde s_\pm^R(n),\delta\tilde s_\pm^I(n),
\delta\tilde{\mathcal E}_\pm^R(n),\delta\tilde{\mathcal E}_\pm^I(n)\rbrace
\ee 
One of the main interests in this appendix is to explicitly compute the UV counter-term $I_{\mathrm{ct}}$ in a UV complete theory. It comes from the high frequency regime, so we focus on the non-zero modes from now on. Their quadratic action is block diagonal: 
\be \label{eq:JT_quad_action}
I^{(2)}_{\mathrm{eff}} = \pi \sum_{n>0} \left(\vec{X}_{+} (n) \mathcal{M}_+(n) \vec{X}_+ (n) + \vec{X}_- (n) \mathcal{M}_-(n) \vec{X}_- (n)\right)
\ee
The matrices $\mathcal{M}_\pm(n)$ are explicitly given by:
\begingroup
\setlength{\arraycolsep}{5pt}
\be\label{eq:jt_matrix_blocks}
\mathcal M_\pm(n)=
\begin{pmatrix}
\lambda^2T F_\pm&0&0& -iL_\pm\\[6pt]
0&\lambda^2T F_\pm& iL_\pm&0\\[6pt]
0& iL_\pm&\lambda^2T F_\pm H&0\\[6pt]
 -iL_\pm &0&0&\lambda^2T F_\pm H
\end{pmatrix}.
\ee
\endgroup
We write the frequency dependence in terms of the dimensionless quantity $\tilde{\omega}_n=\omega_n\sqrt{C/E}$ for $\omega_n=2\pi n/T$. The factors $\lbrace L_\pm,F_\pm,H\rbrace$ are
\bea\label{eq:jt_full_kernels}
\lbrace L_+,L_-\rbrace &=&2n(1+\tilde{\omega}_n^2)\lbrace 1,1\rbrace+\frac{\lambda^2T\tan(\pi d)}{2\pi\tilde{\omega}_n \sqrt{EC}}\lbrace U(\tilde{\omega}_n),B(\tilde{\omega}_n)\rbrace,\nonumber\\
\lbrace F_+,F_-\rbrace &=&\frac{E h(\tilde{\omega}_n)}{\pi C}\cosh^2{\left(\frac{\pi\tilde{\omega}_n}{2}\right)}
\left\lbrace(d-1)^2\tilde{\omega}_n^2,\frac{\tilde{\omega}_n^2+d^2}{2d+1}\right\rbrace,\nonumber\\
H(n)&=&\frac{C}{E^3}\frac{\tanh^2(\pi\tilde{\omega}_n/2)}{\tilde{\omega}_n^2},\;\;\;\;d=2\Delta
\eea
and the intermediate factors are given by:
\bea\label{eq:jt_full_kernels_2}
h(\tilde{\omega})&=&\left(\frac{E}{C}\right)^{d-1/2}\frac{|\Gamma(d+i\tilde{\omega})|^2}{\Gamma(2d)},\nonumber\\
U(\tilde{\omega})&=&(d-1)\tilde{\omega}^2\left[d\,h(0)+(d-1)\cosh(\pi\tilde{\omega})h(\tilde{\omega})\right],\nonumber\\
B(\tilde{\omega})&=&\frac{(\tilde{\omega}^2+d^2)\cosh(\pi\tilde{\omega})h(\tilde{\omega})-d^2h(0)}{2d+1}-d\tilde{\omega}^2h(0)
\eea

We emphasize that the above expressions (\ref{eq:JT_quad_action}), (\ref{eq:jt_full_kernels}), and (\ref{eq:jt_full_kernels_2}) encode the UV-complete frequency dependence of the one-loop determinant in the coupled JT theory and are valid for all $n>0$. Its leading order in the low frequency expansion in small $\tilde{\omega}_n$ gives the collective-coordinate action in the adiabatic approximation derived in section (\ref{sec:two_sphere_hydro}).

Now we can study the UV corrections beyond that and in doing this extract the counter-term $I_{\mathrm{ct}}$ explicitly.  To proceed, we perform the full Gaussian integral for the non-zero fluctuation modes and obtain the determinant:
\bea\label{eq:jt_determinant_ratio}
Z_{\mathrm{fluc}}&=&\int\prod_{n>0,\,\sigma=\pm}4\,d^4\vec X_\sigma(n)\;
 e^{-I_{\mathrm{eff}}^{(2)}}
=\prod_{n>0,\,\sigma=\pm}\frac{4}{\sqrt{\det\mathcal M_\sigma(n)}}\nonumber\\
&=&\prod_{n>0,\,\sigma=\pm}
\frac{4}{L_\sigma(n)^2+
\lambda^4T^2F_\sigma(n)^2H(n)}.
\eea
After plugging in (\ref{eq:jt_full_kernels}) and (\ref{eq:jt_full_kernels_2}), we can decompose the log determinant into:
\bea\label{eq:jt_log_determinant}
\log Z_{\mathrm{fluc}}&=&-4\sum_{n>0} \ln{n}-\sum_{n>0}\log\left(1+\frac{\gamma^2T^2}{4\pi^2 n^2}\right)-4\sum_{n>0}\log\left(1+\tilde{\omega}_n^2\right)\nonumber\\
&&-\frac{\lambda^2 \tan(\pi d)}{E}\sum_{n>0}
\frac{U(\tilde{\omega}_n)+B(\tilde{\omega}_n)}{\tilde{\omega}_n^2(1+\tilde{\omega}_n^2)}+\cdots,
\eea
where the relaxation coefficient is
\be\label{eq:jt_relaxation_rate}
\gamma=\frac{\pi\lambda^2d^2h(0)}{2\sqrt{CE}(2d+1)},
\ee
and ... denotes higher order terms in small $\lambda$. The first two terms are the hydrodynamic results we obtained in section (\ref{sec:two_sphere_hydro}). Via the $\zeta$-regularization they give as before:
\be\label{eq:jt_soft_factor}
Z_{\mathrm{fluc}}^{\mathrm{hydro}}(\lambda)
=\frac{1}{(2\pi)^2}\frac{\gamma T/2}{\sinh(\gamma T/2)}.
\ee
The third term is of order $\mathcal{O}(\lambda^0)$ and is the UV-correction from the JT kinetic term. It can also be evaluated via $\zeta$-regularization to be: 
\be
\prod^{\infty}_{n=1} (1+\tilde{\omega}_n^2)^{-4} = \left[2\sinh{\left(\frac{T}{2}\sqrt{\frac{E}{C}}\right)}\right]^{-4}
\ee
and gives a UV correction $I_{\mathrm{UV}}(0)\propto 2T\sqrt{\frac{E}{C}}$. However, in the hydrodynamic theory of JT gravity, this correction should have already been added at $\lambda=0$. So the counter-term $I_{\mathrm{ct}}$ we define in section (\ref{sec:two_sphere_hydro}) is only the additional UV correction due to finite $\lambda$: $I_{\mathrm{ct}} = I_{\mathrm{UV}}(\lambda)-I_{\mathrm{UV}}(0)$. This is captured by the term in the second line of (\ref{eq:jt_log_determinant}). It can be understood as the UV correction to the adiabatic approximation used in evaluating $\Delta I$. For large $T$, the sum can be replaced by an integral:
\be
I_{\mathrm{ct}} = \frac{\lambda^2 T \tan{(\pi d)}}{2\pi \sqrt{EC}}\int^\infty_0 d\omega \frac{U(\omega)+B(\omega)}{\omega^2(1+\omega^2)}
\ee
This can be evaluated by decomposing the integrand into: 
\be 
\frac{U(\omega)+B(\omega)}{\omega^2(1+\omega^2)} = \frac{A_1\left(\cosh{(\pi \omega)}h(\omega)-h(0)\right)}{\omega^2} + \frac{A_2\cosh{(\pi \omega)}h(\omega)}{1+\omega^2} +  \frac{A_3}{1+\omega^2} 
\ee
with coefficients given by: 
\be
A_1 = \frac{d^2}{2d+1},\;A_2 = \frac{2(d-1)(d^2-d-1)}{2d+1},\;A_3 = \frac{2d(d^2-d-1)}{2d+1}h(0)
\ee
Using the following integral identities: 
\bea
&&\int^\infty_0  d\omega \frac{\cosh{(\pi \omega)}h(\omega)}{1+\omega^2} = - \frac{\pi d h(0)}{2(d-1)},\;\int^\infty_0 \frac{d\omega}{1+\omega^2}=\frac{\pi}{2}\nonumber\\
&& \int^\infty_0 d\omega \frac{\cosh{(\pi \omega)}h(\omega)-h(0)}{\omega^2} = -\frac{\pi^2 h(0)}{2}\cot{(\pi d)}
\eea
we finally obtain that: 
\be 
I_{\mathrm{ct}}= -\frac{\lambda^2 T \pi d^2 h(0)}{4\sqrt{EC}(2d+1)} = -\frac{\gamma T}{2}
\ee
This is indeed the precise form of the counter-term (\ref{eq:ct_proposal}).

\section{Disk one-loop matter correction in the canonical ensemble}\label{app:gaussian_decay}
The one-loop determinant on the complex geometry is given by the matter path integral: 
\be
Z^{\Phi}_{disc} (\beta\pm iT) = \int \mathcal{D}\Phi\,\mathcal{D}\bar\Phi\; e^{-\int\sqrt{g}\,\bar\Phi\left(-\nabla^2_{\beta\pm iT}+m^2\right)\Phi}
\ee
where $\Phi$ denotes a bulk field with mass $m$. The subscript $\beta\pm iT$ for the bulk Laplacian $\nabla^2$ indicates the analytically-continued bulk metric (\ref{eq:renyi_metric}) by $n\to \beta \pm iT$. Formally this gives a functional determinant for the Laplacian in a complex metric. As with the on-shell action, it can be obtained from the Euclidean results by $n\to \beta\pm iT$. So we first do the computation with real modular inverse temperature $n=\beta$. 

To this end, we apply the DHS formula \cite{DHS}, which relates the Euclidean functional determinant in a black hole background to its quasi-normal frequencies: 
\bea \label{eq:DHS}
\int \mathcal{D}\Phi\,\mathcal{D}\bar\Phi\; e^{-\int\sqrt{g}\,\bar\Phi\left(-\nabla^2_\beta+m^2\right)\Phi}
=\mathcal{N}\prod_{\omega,\bar{\omega}} \prod^{\infty}_{n=-\infty} \left(i\omega+\frac{2\pi |n|}{\beta}\right)^{-\frac{N_{\omega}}{2}}\left(i\bar{\omega}-\frac{2\pi |n|}{\beta}\right)^{-\frac{N_{\bar{\omega}}}{2}} 
\eea
where $\mathcal{N}$ includes local terms that can be absorbed into the coupling renormalization. On the right-hand-side of (\ref{eq:DHS}), $\omega,\bar{\omega}$ are the quasi-normal and anti-quasi-normal frequencies respectively, and $N_{\omega,\bar{\omega}}$ are their degeneracies. They are properties of both the black hole background and the nature of the field: 
\be
\omega = \omega(M,m),\;\;\bar{\omega} = \bar{\omega}(M,m)
\ee
where $M=2\pi M_{\mathrm{ADM}}$ is the energy conjugate to modular time $t$, fixed by $\beta$. 

\subsection{Massive scalar}
\begin{figure}[htbp]
\centering
\includegraphics[scale=0.8]{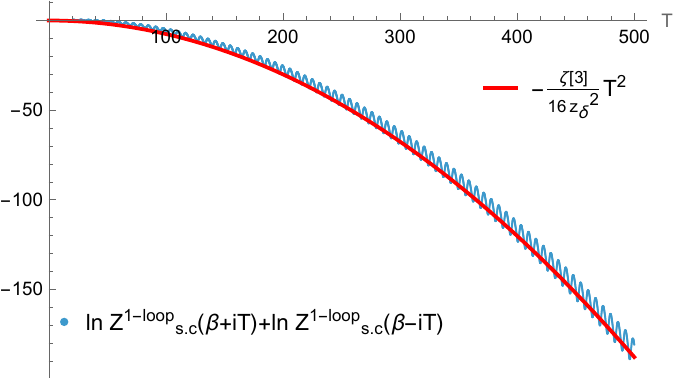}
\caption{A comparison between the approximation (\ref{eq:sc_det_approximate}) and numerical computations based on (\ref{eq:sc_det_3}) for $\beta=2.5,\Delta = 2.4, z_{\delta} = 10$. }\label{fig:single_cone_determinant}
\end{figure}

We illustrate this by studying the one-loop factor from a bulk real scalar field $\Phi$ of mass $m$. Its determinant is the square root of the complex-scalar result (\ref{eq:DHS}). The quasi-normal frequencies can be found by solving the scalar wave equation and imposing an in-falling boundary condition at the horizon. In the entanglement context we have additional boundaries at $z=\pm z_{\delta}$, where the boundary condition for $\Phi$ is the bulk dual of the conformal boundary condition one imposes in the boundary CFT, which we assume to be a Neumann condition: $\partial_z\Phi(r,t,\pm z_{\delta})=0$. Based on this, we work with the ansatz in modular time $t$:
\be 
\Phi(r,t,z)=\phi_{k,\omega}(r) \cos{\left[\frac{k\pi(z+z_{\delta})}{2z_{\delta}}\right]}e^{-i\omega t}
\ee
for integer $k\geq 0$. We then solve the radial equation: 
\be\label{eq:QNM_eq}
\left[\left(r^2-r_h^2\right)\partial^2_r+\left(3r^2-r_h^2\right)r^{-1}\partial_r+\frac{\omega^2}{4\pi^2(r^2-r_h^2)}-\frac{k^2\pi^2 }{4r^2 z_{\delta}^2} -m^2\right] \phi_{k,\omega}(r) =0
\ee
Imposing the in-falling boundary condition gives the solution: 
\be\label{eq:QNM_sol}
\phi(r) =  x^{i\kappa_k/2}(x-1)^{-i\tilde{\omega}/2}
{}_2F_1\left(\tfrac{\Delta+i(\kappa_k-\tilde{\omega})}{2}, \tfrac{2-\Delta+i(\kappa_k-\tilde{\omega})}{2},1-i\tilde{\omega},1-x\right)
\ee
where we have defined $x=r^2/r_h^2$,\;$\kappa_k = k\pi/(2r_h z_{\delta}),\;\tilde{\omega}=\omega/(2\pi r_h),\;m^2 = \Delta(\Delta-2)$. Imposing the normalizable boundary condition at $x\to \infty$ then fixes the quasi-normal and anti-quasi-normal frequencies to take the following quantized values \footnote{We contrast this to the case of a BTZ black hole on a circle \cite{DHS}, which has a further doubling of quasi-normal/anti-quasi-normal frequencies for $k\neq 0$, coming from the two independent spatial harmonics at each nonzero momentum magnitude.}: 
\be\label{eq:Phi_qnm}
\omega_j(k) = 2\pi\left[\pm\frac{k\pi}{2z_{\delta}} -ir_h(\Delta+2j)\right],\;\;\bar{\omega}_j(k)= \omega_j(k)^*,\;\;\;j=0,1,2,...
\ee
Plugging (\ref{eq:Phi_qnm}) into the DHS formula (\ref{eq:DHS}) and performing the analytic continuation $\beta \to \beta \pm iT$, we obtain:
\be \label{eq:sc_det_2}
\ln{{Z^{\Phi}_{disc}}(\beta\pm iT)} = -\frac{1}{2}\sum_{\kappa \geq 0} (\kappa+1) \ln{\left[1-e^{-\frac{4z_{\delta}(\kappa+\Delta)}{\beta\pm iT}}\right]}
\ee
where a divergent local counter-term has been absorbed as in \cite{DHS}. To further re-sum this, we first expand the log and re-sum over $\kappa$ to obtain: \footnote{This is done using the identity: 
\be
\sum_{\kappa\geq 0} (\kappa+1)  \ln{(1-a x^\kappa)} = -\sum_{m\geq 1}\frac{a^m}{m}\sum_{\kappa\geq 0} (\kappa+1)x^{m\kappa} = -\sum_{m\geq 1}\frac{a^m}{m(1-x^m)^2}
\ee}
\be\label{eq:sc_det_3} 
\ln{{Z^{\Phi}_{disc}}(\beta\pm iT)} = \frac{1}{2}\sum_{m\geq 1} \frac{m^{-1}\cdot e^{-\frac{4mz_{\delta}\Delta}{\beta\pm iT}}}{\left(1-e^{-\frac{4mz_{\delta}}{\beta\pm iT}}\right)^2}
\ee
At large $T$ and for $\Delta = \mathcal{O}(1)$, the sum is dominated by those with $m \leq |\frac{\beta\pm iT}{4z_{\delta}}|$, for which we can approximate the summand at large $T$ and obtain: 
\bea\label{eq:sc_det_approximate}
\ln{{Z^{\Phi}_{disc}}(\beta\pm iT)}\approx \frac{(\beta\pm iT)^2}{32 z^2_{\delta}} \sum^{\frac{|\beta+iT|}{4z_{\delta}}}_{m=1} \frac{1}{m^3}
\approx \frac{\zeta(3)}{32z_{\delta}^2} (\beta\pm iT)^2
\eea
where in the last step we have approximated the upper limit of the sum by $\infty$. The two disc factors enter the SFF as a product. Combining their large-$T$ contributions gives a Gaussian decay:
\be\label{eq:disc_gaussian_decay}
Z^{\Phi}_{dis}(\beta,T)={Z^{\Phi}_{disc}}(\beta+ iT){Z^{\Phi}_{disc}}(\beta- iT) \propto \exp{\left[-\frac{\zeta(3)}{16 z_{\delta}^2} T^2\right]}
\ee
See Figure (\ref{fig:single_cone_determinant}) for a numerical verification of the result. We recall that in a canonical definition of $F_A(\beta,T)$, this Gaussian decaying contribution could be replaced at large $T$ by another disconnected contribution corresponding to the analog of the ``thermal AdS'' geometry in the entanglement context \cite{ChuLiWang:WIP}, which is expected to reveal the IR end of the entanglement spectrum.

To convert the results back into the micro-canonical version, one can simply replace the $\beta$ in (\ref{eq:sc_det_2}) with the saddle-point values at the leading order in small $G_N$:
\be
\beta^*_\pm  = \beta(E) \mp iT
\ee 
according to (\ref{eq:mass_temperature}). By doing this, the $T$-dependence disappears from the one-loop factor, and the micro-canonical determinant is given by:
\bea\label{eq:scalar_micro_expansion}
Z^{\Phi}_{disc}(E)
&=&\exp{\left(-\frac12\sum_{\kappa\geq0}(\kappa+1)\ln\left[1-\widetilde{q}_{E}^{\kappa+\Delta}\right]\right)}\nonumber\\
&=&1+\frac12\widetilde q_{E}^{\Delta}+\cdots,\qquad
\widetilde q_{E}=e^{-4z_\delta/\beta(E)} 
\eea
At fixed $\beta(E)$, the excited corrections are powers of $\widetilde{q}_{E}\propto \delta^{4/\beta(E)}$ and are suppressed by the UV entanglement cut-off $\delta \ll 1$.

\subsection{Boundary graviton}
A special sector in $\text{AdS}_3/\text{CFT}_2$ involves the boundary gravitons, whose excitations are dual to Virasoro descendants. To extract its contribution as a bulk determinant, we invoke the representation for the unnormalized $\mathcal Z_A(\beta)$ as a BCFT partition function on a cylinder with modular parameter $\tau_{\mathrm{disc}} = i\pi\beta/(2z_{\delta})$ that satisfies the open-closed duality:
\be 
\mathcal Z_A(\beta) = Z^{ab}_{\text{open}}(\tau_{\mathrm{disc}}) = Z^{ab}_{\text{closed}}(-1/\tau_{\mathrm{disc}})
\ee
The labels $a,b$ represent the conformal boundary conditions we impose on the entanglement cut-off surface at $z=\pm z_{\delta}$. In the high temperature limit $|\tau_{\mathrm{disc}}|\ll 1$, the closed channel is dominated by the vacuum character:
\be\label{eq:vac_dom_1}
\mathcal Z_A(\beta) \approx G_{BB}\;e^{-2\pi i\Delta_0/\tau_{\mathrm{disc}}}\prod_{n\geq2}\left(1-e^{-2\pi in/\tau_{\mathrm{disc}}}\right)^{-1}
\ee
where $G_{BB}=\langle B|I\rangle\langle I|B\rangle\approx 1$, when $|B\rangle$ is the boundary dual of the tensionless brane. The first exponential is identified with the classical contribution with $\Delta_0=-c/24$ being the Casimir energy in the closed channel. Using $\eta(-1/\tau_{\mathrm{disc}})=\sqrt{-i\tau_{\mathrm{disc}}}\,\eta(\tau_{\mathrm{disc}})$ we can decompose the one-loop corrected holographic result into three factors:
\bea\label{eq:vac_dom}
\mathcal Z_A(\beta)&\approx & e^{-2\pi i \Delta_0 /\tau_{\mathrm{disc}}}\;e^{-2\pi i/(24\tau_{\mathrm{disc}})}(1-e^{-2\pi i/\tau_{\mathrm{disc}}})\left(-i\tau_{\mathrm{disc}}\right)^{-1/2}\eta\left(\tau_{\mathrm{disc}}\right)^{-1}\nonumber\\
&=& (\text{classical})\times Z^{\mathrm{zm}}_{disc}(\beta)\times Z^{\mathrm{vir}}_{disc}(\beta)
\eea
In particular, $(\text{classical})=e^{-2\pi i \Delta_0/\tau_{\mathrm{disc}}}$ is the bulk on-shell action (\ref{eq:renyi_action}); $Z^{\mathrm{zm}}_{disc}(\beta)=(1-e^{-2\pi i /\tau_{\mathrm{disc}}}) (-i\tau_{\mathrm{disc}})^{-1/2}$ represents contributions from the zero-mode sector associated with the $SL(2,R)$ invariance of the closed-channel vacuum state; and finally $Z^{\mathrm{vir}}_{disc}(\beta)=e^{-2\pi i/(24\tau_{\mathrm{disc}})}\eta(\tau_{\mathrm{disc}})^{-1}$
is identified with the one-loop correction from the boundary graviton excitations.

We can perform the analytic continuation $\beta \to \beta \pm iT$ for each factor in (\ref{eq:vac_dom}). By doing so we obtain that for the disconnected saddle-points, the zero-mode factor decays as a power-law at large $T$: 
\be
\lim_{T\to \infty} \left(1-e^{-\frac{4z_{\delta}}{\beta \pm iT}}\right) \sqrt{\frac{2z_{\delta}}{\pi(\beta\pm iT)}} \propto T^{-3/2}
\ee
This can be identified as the counterpart in 2d CFT of the Schwarzian contribution in JT gravity \cite{Maldacena:2016upp}. The graviton one-loop factor contains an oscillatory product in $T$: 
\be 
Z^{\mathrm{vir}}_{disc}(\beta \pm iT) = e^{-\frac{z_\delta}{6(\beta\pm iT)}+\frac{\pi^2(\beta\pm iT)}{24z_\delta}}\prod_{n\geq 1}\left[1-e^{-\frac{n\pi^2(\beta \pm iT)}{z_{\delta}}}\right]^{-1}
\ee
Despite being localized on the asymptotic boundary, one can formally identify the quasi-normal frequencies of the boundary gravitons through (\ref{eq:DHS}) as: 
\be\label{eq:graviton_qnm}
\omega_n = \bar{\omega}_n = \frac{n\pi^2}{z_{\delta}},\;\;n = 1,2,...
\ee
They are real normal frequencies, and correspond to one copy (as a result of the boundary condition) of the Virasoro descendant modes $\lbrace \mathcal{L}_{-n}\rbrace_{n\geq 1}$, universally present in heavy Virasoro characters. Their presence is a special feature of working with $\text{AdS}_3/\text{CFT}_2$.

Similar to the massive bulk scalar, the micro-canonical result is obtained by replacing $\beta\pm iT$ with $\beta(E)$. Returning to the closed-channel product in (\ref{eq:vac_dom_1}), the full gravitational one-loop correction is:
\be
Z^{\mathrm{grav}}_{disc}(E)
=\prod_{n\geq2}\left(1-\widetilde q_{E}^n\right)^{-1}
=1+\widetilde q_{E}^2+...
\ee
The excited corrections are suppressed by small $\delta$ similar to $Z^{\Phi}_{disc}(E)$ in (\ref{eq:scalar_micro_expansion}).

In summary, we identify the following large-$T$ behavior of the disconnected contribution to the canonical result:
\bea
Z_{disc}(\beta+iT)Z_{disc}(\beta-iT)
&\sim & e^{-2a\beta+\frac{z_\delta\beta}{2G_NT^2}+\frac{\pi^2\beta}{12z_\delta}-\sum_{\Phi\in\mathrm{EFT}} \frac{\zeta(3)T^2}{16z_\delta^2}}
\nonumber\\
&\times & T^{-3}\prod_{n\geq1}\left|1-e^{-\frac{n\pi^2(\beta+iT)}{z_\delta}}\right|^{-2},
\eea

\section{Deriving Eq. (\ref{eq:mod_OPE}) from density matrix OPE}\label{app:descendants}

\begin{figure}[htbp]
\centering
\includegraphics[width=\textwidth]{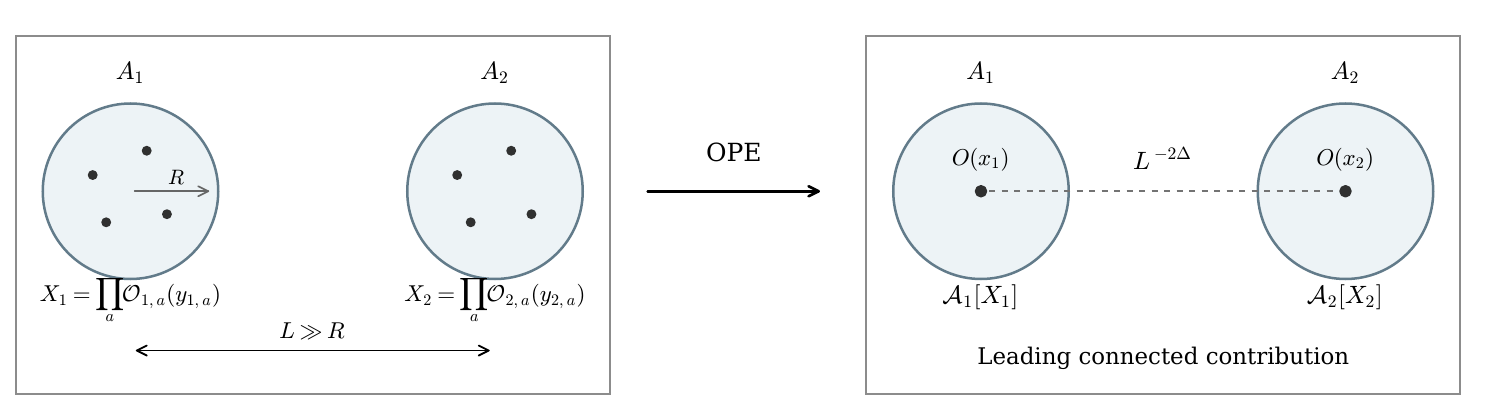}
\caption{OPE clusters $X_1$ and $X_2$ into local operators at $x_1$ and $x_2$.}
\label{fig:cluster_ope}
\end{figure}

In this appendix, we derive the leading order term (\ref{eq:mod_OPE}). Consider two balls $A_i$ of radius $R$ centered at $x_i$, with $|x_1-x_2|=L\gg R$, in the CFT vacuum. At infinite separation their reduced density matrix factorizes. At large but finite separation, we expand about this uncorrelated state:
\be
\rho_{A_1\cup A_2}=\rho_0+\delta\rho+\cdots,\qquad
\rho_0=\rho_{A_1}\otimes\rho_{A_2}.
\ee
The correction $\delta\rho$ encodes correlations between the regions, while each single-region density matrix remains fixed. We determine it by matching expectation values: products of clusters $X_1\otimes X_2$, where $\lbrace X_i \rbrace =\lbrace \prod_{a=1}^{N_i}\mathcal O_{i,a}(y_{i,a}),\;y_{i,a}\in A_i\rbrace$, span the regulated two-region operator algebra, so their connected correlators determine $\delta\rho$. We first compute these correlators by a large-separation OPE, from which $\delta\rho$ can be reconstructed, and then expand $-\log\rho_{A_1\cup A_2}$ to obtain the modular Hamiltonian. The matching condition is:
\be
\operatorname{Tr}\!\left[\delta\rho\,(X_1\otimes X_2)\right]
\simeq\langle X_1X_2\rangle_c=\langle X_1X_2\rangle-\langle X_1\rangle\langle X_2\rangle
\ee
We evaluate the right-hand side by replacing each cluster by local operators at $x_i$, see Figure (\ref{fig:cluster_ope}):
\be
X_i\sim\langle X_i\rangle\mathbf{1}
+\mathcal A_i[X_i]O(x_i)
+\text{descendants}+\text{other primaries}
\ee
Here we consider a scalar primary of conformal dimension $\Delta$. Its OPE coefficient $\mathcal{A}_i[X_i]$ is extracted by evaluating:
\be
\mathcal A_i[X_i]=\lim_{|y|\to\infty}|y|^{2\Delta}\langle X_iO(x_i+y)\rangle.
\ee
Its contribution to the connected correlator is given by:
\be
\langle X_1X_2\rangle_c\simeq
\mathcal A_1[X_1]\mathcal A_2[X_2]\langle O(x_1)O(x_2)\rangle
\simeq \frac{\mathcal A_1[X_1]\mathcal A_2[X_2]}{L^{2\Delta}}.
\ee

In the $|y|\to \infty$ limit, the operator $O(x_i+y)$ is proportional to the modular reflection of $O(x_i)$ using $\rho_{A_i}$. Writing the hyperbolic-frame operator as $O_{A_i}=(R/2)^\Delta O(x_i)$ including the conformal factor gives:
\be
\mathcal A_i[X_i]=(2R)^\Delta\operatorname{Tr}\!\left(\rho_{A_i}^{1/2} O_{A_i}\rho_{A_i}^{1/2}X_i\right).
\ee
Matching these expectation values:
\be\label{eq:ope_density_matrix}
\operatorname{Tr}\!\left[\delta\rho\,(X_1\otimes X_2)\right] = \lambda \operatorname{Tr}\!\left[\rho^{1/2}_0 O_{A_1}\otimes O_{A_2}\rho^{1/2}_0 \left(X_1 \otimes X_2\right)\right] + ...
\ee
for an arbitrary cluster pair $X_1\otimes X_2$ identifies the leading density-matrix correction:
\be 
\delta \rho = \lambda \rho^{1/2}_0 O_{A_1}\otimes  O_{A_2} \rho_0^{1/2}+...
\ee
where $\lambda = (2R/L)^{2\Delta}$.

We now expand the modular Hamiltonian $\Delta K_{A_1,A_2}=\log\rho_0-\log(\rho_0+\delta\rho)$. In the modular basis $\rho_0|a\rangle=e^{-k_a}|a\rangle$, its leading-order matrix elements are
\be
\langle a|\Delta K_{A_1,A_2}|b\rangle
=\frac{-\lambda(k_a-k_b)}{2\sinh[(k_a-k_b)/2]}
\langle a| O_{A_1}\otimes O_{A_2}|b\rangle,
\ee
This can be rewritten into the modular-flow integral form of (\ref{eq:mod_OPE}):
\be
\Delta K_{A_1,A_2}=-\lambda\int_{-\infty}^{\infty}\frac{\pi\;ds}{2\cosh^2(\pi s)}
\left(e^{iK_{A_1}s} O_{A_1}e^{-iK_{A_1}s}\right)
\left(e^{iK_{A_2}s} O_{A_2}e^{-iK_{A_2}s}\right),
\ee
by using the identity:
\be
\frac{\omega}{2\sinh(\omega/2)}
=\int_{-\infty}^{\infty}\frac{\pi\;e^{i\omega s} ds}{2\cosh^2(\pi s)}.
\ee

\bibliographystyle{JHEP}
\bibliography{ref}

@article{Maldacena:1997re,
    author = "Maldacena, Juan Martin",
    title = "{The Large $N$ limit of superconformal field theories and supergravity}",
    eprint = "hep-th/9711200",
    archivePrefix = "arXiv",
    reportNumber = "HUTP-97-A097, HUTP-98-A097",
    doi = "10.4310/ATMP.1998.v2.n2.a1",
    journal = "Adv. Theor. Math. Phys.",
    volume = "2",
    pages = "231--252",
    year = "1998"
}

@article{Penington:2019kki,
    author = "Penington, Geoff and Shenker, Stephen H. and Stanford, Douglas and Yang, Zhenbin",
    title = "{Replica wormholes and the black hole interior}",
    eprint = "1911.11977",
    archivePrefix = "arXiv",
    primaryClass = "hep-th",
    doi = "10.1007/JHEP03(2022)205",
    journal = "JHEP",
    volume = "03",
    pages = "205",
    year = "2022"
}

@article{winer2022hydrodynamic,
  title={Hydrodynamic theory of the connected spectral form factor},
  author={Winer, Michael and Swingle, Brian},
  journal={Physical Review X},
  volume={12},
  number={2},
  pages={021009},
  year={2022},
  publisher={APS}
}

@article{Fujita:2011fp,
    author = "Fujita, Mitsutoshi and Takayanagi, Tadashi and Tonni, Erik",
    title = "{Aspects of AdS/BCFT}",
    eprint = "1108.5152",
    archivePrefix = "arXiv",
    primaryClass = "hep-th",
    reportNumber = "IPMU-11-0136, MIT-CTP-4289",
    doi = "10.1007/JHEP11(2011)043",
    journal = "JHEP",
    volume = "11",
    pages = "043",
    year = "2011"
}

@article{Jafferis:2025yxt,
    author = "Jafferis, Daniel L. and Rozenberg, Liza and Wang, Diandian",
    title = "{Open-closed 3d gravity as a random ensemble}",
    eprint = "2506.19817",
    archivePrefix = "arXiv",
    primaryClass = "hep-th",
    doi = "10.1007/JHEP10(2025)228",
    journal = "JHEP",
    volume = "10",
    pages = "228",
    year = "2025"
}

@article{Maldacena:2015waa,
    author = "Maldacena, Juan and Shenker, Stephen H. and Stanford, Douglas",
    title = "{A bound on chaos}",
    eprint = "1503.01409",
    archivePrefix = "arXiv",
    primaryClass = "hep-th",
    doi = "10.1007/JHEP08(2016)106",
    journal = "JHEP",
    volume = "08",
    pages = "106",
    year = "2016"
}

@article{Shenker:2013pqa,
    author = "Shenker, Stephen H. and Stanford, Douglas",
    title = "{Black holes and the butterfly effect}",
    eprint = "1306.0622",
    archivePrefix = "arXiv",
    primaryClass = "hep-th",
    reportNumber = "SU-ITP-13-08",
    doi = "10.1007/JHEP03(2014)067",
    journal = "JHEP",
    volume = "03",
    pages = "067",
    year = "2014"
}

@article{Saad:2018bqo,
    author = "Saad, Phil and Shenker, Stephen H. and Stanford, Douglas",
    title = "{A semiclassical ramp in SYK and in gravity}",
    eprint = "1806.06840",
    archivePrefix = "arXiv",
    primaryClass = "hep-th",
    month = "6",
    year = "2018"
}

@article{Ryu:2006bv,
    author = "Ryu, Shinsei and Takayanagi, Tadashi",
    title = "{Holographic derivation of entanglement entropy from AdS/CFT}",
    eprint = "hep-th/0603001",
    archivePrefix = "arXiv",
    reportNumber = "NSF-KITP-06-11, NSF-KITP-06-11",
    doi = "10.1103/PhysRevLett.96.181602",
    journal = "Phys. Rev. Lett.",
    volume = "96",
    pages = "181602",
    year = "2006"
}

@article{Faulkner:2013ana,
    author = "Faulkner, Thomas and Lewkowycz, Aitor and Maldacena, Juan",
    title = "{Quantum corrections to holographic entanglement entropy}",
    eprint = "1307.2892",
    archivePrefix = "arXiv",
    primaryClass = "hep-th",
    doi = "10.1007/JHEP11(2013)074",
    journal = "JHEP",
    volume = "11",
    pages = "074",
    year = "2013"
}

@article{Witten:2018zxz,
    author = "Witten, Edward",
    title = "{APS Medal for Exceptional Achievement in Research: Invited article on entanglement properties of quantum field theory}",
    eprint = "1803.04993",
    archivePrefix = "arXiv",
    primaryClass = "hep-th",
    doi = "10.1103/RevModPhys.90.045003",
    journal = "Rev. Mod. Phys.",
    volume = "90",
    number = "4",
    pages = "045003",
    year = "2018"
}

@article{Casini:2011kv,
    author = "Casini, Horacio and Huerta, Marina and Myers, Robert C.",
    title = "{Towards a derivation of holographic entanglement entropy}",
    eprint = "1102.0440",
    archivePrefix = "arXiv",
    primaryClass = "hep-th",
    doi = "10.1007/JHEP05(2011)036",
    journal = "JHEP",
    volume = "05",
    pages = "036",
    year = "2011"
}

@article{Chen:2022eyi,
    author = "Chen, Liangyu and Wang, Huajia",
    title = "{Causal shadow and non-local modular flow: from degeneracy to perturbative genesis by correlation}",
    eprint = "2211.12064",
    archivePrefix = "arXiv",
    primaryClass = "hep-th",
    doi = "10.1007/JHEP02(2023)052",
    journal = "JHEP",
    volume = "02",
    pages = "052",
    year = "2023"
}

@article{Cardy:1989ir,
    author = "Cardy, John L.",
    title = "{Boundary Conditions, Fusion Rules and the Verlinde Formula}",
    journal = "Nucl. Phys. B",
    volume = "324",
    pages = "581--596",
    year = "1989",
    doi = "10.1016/0550-3213(89)90521-X"
}

@article{Hung:2019bnq,
    author = "Hung, Ling Yan and Wong, Gabriel",
    title = "{Entanglement branes and factorization in conformal field theory}",
    eprint = "1912.11201",
    archivePrefix = "arXiv",
    primaryClass = "hep-th",
    doi = "10.1103/PhysRevD.104.026012",
    journal = "Phys. Rev. D",
    volume = "104",
    number = "2",
    pages = "026012",
    year = "2021"
}

@article{DHS,
    author = "Denef, Frederik and Hartnoll, Sean A. and Sachdev, Subir",
    title = "{Black hole determinants and quasinormal modes}",
    eprint = "0908.2657",
    archivePrefix = "arXiv",
    primaryClass = "hep-th",
    reportNumber = "NSF-KITP-09-145",
    doi = "10.1088/0264-9381/27/12/125001",
    journal = "Class. Quant. Grav.",
    volume = "27",
    pages = "125001",
    year = "2010"
}

@misc{ChuLiWang:WIP,
  author       = {Chu, J. and Li, X. and Wang, H.},
  howpublished = {Work in progress}
}

@article{Maldacena:2016upp,
  author        = {Maldacena, Juan and Stanford, Douglas and Yang, Zhenbin},
  title         = {Conformal symmetry and its breaking in two dimensional
                   Nearly Anti-de-Sitter space},
  journal       = {PTEP},
  volume        = {2016},
  pages         = {12C104},
  year          = {2016},
  doi           = {10.1093/ptep/ptw124},
  eprint        = {1606.01857},
  archivePrefix = {arXiv},
  primaryClass  = {hep-th}
}

@article{JLMS,
    author = "Jafferis, Daniel L. and Lewkowycz, Aitor and Maldacena, Juan and Suh, S. Josephine",
    title = "{Relative entropy equals bulk relative entropy}",
    eprint = "1512.06431",
    archivePrefix = "arXiv",
    primaryClass = "hep-th",
    reportNumber = "NSF-KITP-15-162, NSF-KITP-15-162",
    doi = "10.1007/JHEP06(2016)004",
    journal = "JHEP",
    volume = "06",
    pages = "004",
    year = "2016"
}

@article{Faulkner:2018faa,
    author = "Faulkner, Thomas and Li, Min and Wang, Huajia",
    title = "{A modular toolkit for bulk reconstruction}",
    eprint = "1806.10560",
    archivePrefix = "arXiv",
    primaryClass = "hep-th",
    doi = "10.1007/JHEP04(2019)119",
    journal = "JHEP",
    volume = "04",
    pages = "119",
    year = "2019"
}

@article{Jan:2020,
    author = "De Boer, Jan and Lamprou, Lampros",
    title = "{Holographic Order from Modular Chaos}",
    eprint = "1912.02810",
    archivePrefix = "arXiv",
    primaryClass = "hep-th",
    doi = "10.1007/JHEP06(2020)024",
    journal = "JHEP",
    volume = "06",
    pages = "024",
    year = "2020"
}

@article{ChenLudwig,
    author = "Chen, Xiao and Ludwig, Andreas W. W.",
    title = "{Universal Spectral Correlations in the Chaotic Wave Function, and the Development of Quantum Chaos}",
    eprint = "1710.02686",
    archivePrefix = "arXiv",
    primaryClass = "cond-mat.str-el",
    doi = "10.1103/PhysRevB.98.064309",
    journal = "Phys. Rev. B",
    volume = "98",
    number = "6",
    pages = "064309",
    year = "2018"
}

@article{ChangEtAl,
    author = "Chang, Po-Yao and Chen, Xiao and Gopalakrishnan, Sarang and Pixley, J. H.",
    title = "{Evolution of Entanglement Spectra under Generic Quantum Dynamics}",
    eprint = "1811.00029",
    archivePrefix = "arXiv",
    primaryClass = "cond-mat.dis-nn",
    doi = "10.1103/PhysRevLett.123.190602",
    journal = "Phys. Rev. Lett.",
    volume = "123",
    number = "19",
    pages = "190602",
    year = "2019"
}

@article{CaoFaulkner,
    author = "Cao, Xuchen and Faulkner, Thomas",
    title = "{Ramp from replica trick}",
    eprint = "2405.15873",
    archivePrefix = "arXiv",
    primaryClass = "hep-th",
    doi = "10.1007/JHEP01(2025)104",
    journal = "JHEP",
    volume = "01",
    pages = "104",
    year = "2025"
}

@article{ChandrasekaranFaulknerLevine,
    author = "Chandrasekaran, Venkatesa and Faulkner, Thomas and Levine, Adam",
    title = "{Scattering strings off quantum extremal surfaces}",
    eprint = "2108.01093",
    archivePrefix = "arXiv",
    primaryClass = "hep-th",
    doi = "10.1007/JHEP08(2022)143",
    journal = "JHEP",
    volume = "08",
    pages = "143",
    year = "2022"
}

@article{deBoerEtAl,
    author = "de Boer, Jan and Najian, Bahman and van der Heijden, Jeremy and Zukowski, Claire",
    title = "{Modular chaos, operator algebras, and the Berry phase}",
    eprint = "2505.04682",
    archivePrefix = "arXiv",
    primaryClass = "hep-th",
    doi = "10.1007/JHEP09(2025)086",
    journal = "JHEP",
    volume = "09",
    pages = "086",
    year = "2025"
}

@article{Bekenstein:1973ur,
    author = "Bekenstein, Jacob D.",
    title = "{Black holes and entropy}",
    doi = "10.1103/PhysRevD.7.2333",
    journal = "Phys. Rev. D",
    volume = "7",
    pages = "2333--2346",
    year = "1973"
}

@article{Hawking:1974sw,
    author = "Hawking, S. W.",
    editor = "Gibbons, G. W. and Hawking, S. W.",
    title = "{Particle Creation by Black Holes}",
    doi = "10.1007/BF02345020",
    journal = "Commun. Math. Phys.",
    volume = "43",
    pages = "199--220",
    year = "1975",
    note = "[Erratum: Commun.Math.Phys. 46, 206 (1976)]"
}

@article{Saad:2019lba,
    author = "Saad, Phil and Shenker, Stephen H. and Stanford, Douglas",
    title = "{JT gravity as a matrix integral}",
    eprint = "1903.11115",
    archivePrefix = "arXiv",
    primaryClass = "hep-th",
    month = "3",
    year = "2019"
}

@article{Chen:2023dc,
    author = "Chen, Yiming and Ivo, Victor and Maldacena, Juan",
    title = "{Comments on the double cone wormhole}",
    eprint = "2310.11617",
    archivePrefix = "arXiv",
    primaryClass = "hep-th",
    doi = "10.1007/JHEP04(2024)124",
    journal = "JHEP",
    volume = "04",
    pages = "124",
    year = "2024"
}

@article{Gubser:1998bc,
    author = "Gubser, S. S. and Klebanov, Igor R. and Polyakov, Alexander M.",
    title = "{Gauge theory correlators from noncritical string theory}",
    eprint = "hep-th/9802109",
    archivePrefix = "arXiv",
    reportNumber = "PUPT-1767",
    doi = "10.1016/S0370-2693(98)00377-3",
    journal = "Phys. Lett. B",
    volume = "428",
    pages = "105--114",
    year = "1998"
}

@article{Witten:1998qj,
    author = "Witten, Edward",
    title = "{Anti de Sitter space and holography}",
    eprint = "hep-th/9802150",
    archivePrefix = "arXiv",
    reportNumber = "IASSNS-HEP-98-15",
    doi = "10.4310/ATMP.1998.v2.n2.a2",
    journal = "Adv. Theor. Math. Phys.",
    volume = "2",
    pages = "253--291",
    year = "1998"
}

@article{Ryu:2006ef,
    author = "Ryu, Shinsei and Takayanagi, Tadashi",
    title = "{Aspects of Holographic Entanglement Entropy}",
    eprint = "hep-th/0605073",
    archivePrefix = "arXiv",
    reportNumber = "NSF-KITP-06-31, KUNS-2021, NSF-KITP-06-31, KUNS-2021",
    doi = "10.1088/1126-6708/2006/08/045",
    journal = "JHEP",
    volume = "08",
    pages = "045",
    year = "2006"
}

@article{Yan:2023rjh,
    author = "Yan, Cynthia",
    title = "{More on torus wormholes in 3d gravity}",
    eprint = "2305.10494",
    archivePrefix = "arXiv",
    primaryClass = "hep-th",
    doi = "10.1007/JHEP11(2023)039",
    journal = "JHEP",
    volume = "11",
    pages = "039",
    year = "2023"
}

@article{Jafferis:2025jle,
    author = "Jafferis, Daniel L. and Rozenberg, Liza and Sarkar, Debmalya and Wang, Diandian",
    title = "{On random matrix statistics of 3d gravity}",
    eprint = "2512.05045",
    archivePrefix = "arXiv",
    primaryClass = "hep-th",
    doi = "10.1007/JHEP08(2026)014",
    journal = "JHEP",
    volume = "08",
    pages = "014",
    year = "2026"
}

@article{DuttaFaulkner,
    author = "Dutta, Souvik and Faulkner, Thomas",
    title = "{A canonical purification for the entanglement wedge cross-section}",
    eprint = "1905.00577",
    archivePrefix = "arXiv",
    primaryClass = "hep-th",
    doi = "10.1007/JHEP03(2021)178",
    journal = "JHEP",
    volume = "03",
    pages = "178",
    year = "2021"
}

@article{Maldacena:2018lmt,
  author = {Maldacena, Juan and Qi, Xiao-Liang},
  title = {Eternal traversable wormhole},
  eprint = {1804.00491},
  archivePrefix = {arXiv},
  primaryClass = {hep-th},
  year = {2018}
}

@article{Hayden:2007cs,
    author = "Hayden, Patrick and Preskill, John",
    title = "{Black holes as mirrors: Quantum information in random subsystems}",
    eprint = "0708.4025",
    archivePrefix = "arXiv",
    primaryClass = "hep-th",
    doi = "10.1088/1126-6708/2007/09/120",
    journal = "JHEP",
    volume = "09",
    pages = "120",
    year = "2007"
}

@article{Sekino:2008he,
    author = "Sekino, Yasuhiro and Susskind, Leonard",
    title = "{Fast Scramblers}",
    eprint = "0808.2096",
    archivePrefix = "arXiv",
    primaryClass = "hep-th",
    doi = "10.1088/1126-6708/2008/10/065",
    journal = "JHEP",
    volume = "10",
    pages = "065",
    year = "2008"
}

@article{Cotler:2016fpe,
    author = "Cotler, Jordan S. and Gur-Ari, Guy and Hanada, Masanori and Polchinski, Joseph and Saad, Phil and Shenker, Stephen H. and Stanford, Douglas and Streicher, Alexandre and Tezuka, Masaki",
    title = "{Black Holes and Random Matrices}",
    eprint = "1611.04650",
    archivePrefix = "arXiv",
    primaryClass = "hep-th",
    doi = "10.1007/JHEP05(2017)118",
    journal = "JHEP",
    volume = "05",
    pages = "118",
    year = "2017",
    note = "[Erratum: JHEP 09, 002 (2018)]"
}

@article{Lewkowycz:2013nqa,
    author = "Lewkowycz, Aitor and Maldacena, Juan",
    title = "{Generalized gravitational entropy}",
    eprint = "1304.4926",
    archivePrefix = "arXiv",
    primaryClass = "hep-th",
    doi = "10.1007/JHEP08(2013)090",
    journal = "JHEP",
    volume = "08",
    pages = "090",
    year = "2013"
}

@article{Bousso:2020yxi,
    author = "Bousso, Raphael and Chandrasekaran, Venkatesa and Rath, Pratik and Shahbazi-Moghaddam, Arvin",
    title = "{Gravity dual of Connes cocycle flow}",
    eprint = "2007.00230",
    archivePrefix = "arXiv",
    primaryClass = "hep-th",
    doi = "10.1103/PhysRevD.102.066008",
    journal = "Phys. Rev. D",
    volume = "102",
    number = "6",
    pages = "066008",
    year = "2020"
}

@article{Kaplan:2022orm,
    author = "Kaplan, Molly and Marolf, Donald",
    title = "{The action of HRT-areas as operators in semiclassical gravity}",
    eprint = "2203.04270",
    archivePrefix = "arXiv",
    primaryClass = "hep-th",
    doi = "10.1007/JHEP08(2022)102",
    journal = "JHEP",
    volume = "08",
    pages = "102",
    year = "2022"
}

@article{Yang:2025lme,
    author = "Yang, Zhenbin and Zhang, Yuzhen and Zheng, Wenwen",
    title = "{Remarks on the de Sitter double cone}",
    eprint = "2505.08647",
    archivePrefix = "arXiv",
    primaryClass = "hep-th",
    reportNumber = "USTC-ICTS/PCFT-25-58",
    doi = "10.1103/jync-pmnz",
    journal = "Phys. Rev. D",
    volume = "113",
    number = "12",
    pages = "126012",
    year = "2026"
}

@article{Ouseph:2023juq,
    author = "Ouseph, Shoy and Furuya, Keiichiro and Lashkari, Nima and Leung, Kwing Lam and Moosa, Mudassir",
    title = "{Local Poincar{\'e} algebra from quantum chaos}",
    eprint = "2310.13736",
    archivePrefix = "arXiv",
    primaryClass = "hep-th",
    doi = "10.1007/JHEP01(2024)112",
    journal = "JHEP",
    volume = "01",
    pages = "112",
    year = "2024"
}

@article{Hung:2011nu,
    author = "Hung, Ling-Yan and Myers, Robert C. and Smolkin, Michael and Yale, Alexandre",
    title = "Holographic Calculations of Renyi Entropy",
    eprint = "1110.1084",
    archivePrefix = "arXiv",
    primaryClass = "hep-th",
    doi = "10.1007/JHEP12(2011)047",
    journal = "JHEP",
    volume = "12",
    pages = "047",
    year = "2011"
}

@article{Calabrese:2010he,
    author = "Calabrese, Pasquale and Cardy, John and Tonni, Erik",
    title = "{Entanglement entropy of two disjoint intervals in conformal field theory II}",
    eprint = "1011.5482",
    archivePrefix = "arXiv",
    primaryClass = "hep-th",
    reportNumber = "NSF-KITP-10-152, MIT-CTP 4194",
    doi = "10.1088/1742-5468/2011/01/P01021",
    journal = "J. Stat. Mech.",
    volume = "1101",
    pages = "P01021",
    year = "2011"
}

@article{Cardy:2013nua,
    author = "Cardy, John",
    title = "{Some results on the mutual information of disjoint regions in higher dimensions}",
    eprint = "1304.7985",
    archivePrefix = "arXiv",
    primaryClass = "hep-th",
    doi = "10.1088/1751-8113/46/28/285402",
    journal = "J. Phys. A",
    volume = "46",
    pages = "285402",
    year = "2013"
}

@article{Agon:2015ftl,
    author = "Ag{\'o}n, Cesar and Faulkner, Thomas",
    title = "{Quantum Corrections to Holographic Mutual Information}",
    eprint = "1511.07462",
    archivePrefix = "arXiv",
    primaryClass = "hep-th",
    doi = "10.1007/JHEP08(2016)118",
    journal = "JHEP",
    volume = "08",
    pages = "118",
    year = "2016"
}

@article{Chen:2022cmr,
    author = "Chen, Liangyu and Wang, Huajia",
    title = "{Shape dependence of mutual information in the OPE limit: linear responses}",
    eprint = "2207.05268",
    archivePrefix = "arXiv",
    primaryClass = "hep-th",
    doi = "10.1007/JHEP10(2022)101",
    journal = "JHEP",
    volume = "10",
    pages = "101",
    year = "2022"
}

@article{Jia:2026uuz,
    author = "Jia, Hewei Frederic",
    title = "{Mutual information to all orders from modular flow: a replica-free approach}",
    eprint = "2611.03630",
    archivePrefix = "arXiv",
    primaryClass = "hep-th",
    month = "11",
    year = "2026"
}

\end{document}